\documentclass[a4paper,11pt]{article}

\usepackage{jinstpub} 
\usepackage{booktabs} 
\usepackage{lineno}
\usepackage{graphicx} 
\usepackage{subfig}
\usepackage{float}
\usepackage{lscape}

\graphicspath{{./pics/}}  

\title{Design, construction and commissioning of the PandaX-30T liquid xenon management system}
\author[a]{Xiuli Wang,}
\author[a]{Zhuoqun Lei,}
\author[a,1]{Yonglin Ju,\note{Corresponding author.}}
\author[b,c,d]{Jianglai Liu,}
\author[b,d]{Ning Zhou,}
\author[e]{Yu Chen,}
\author[b,c,d]{Zhou Wang,}
\author[c]{Xiangyi Cui,}
\author[b,d]{Yue Meng,}
\author[b,d]{Li Zhao}

\affiliation[a]{Institute of Refrigeration and Cryogenics, Shanghai Jiao Tong University, Shanghai 200240, China}
\affiliation[b]{INPAC, School of Physics and Astronomy, Shanghai Jiao Tong University, Shanghai Key Laboratory for Particle Physics and Cosmology, Key Laboratory for Particle Astrophysics and Cosmology (MOE), \\Shanghai 200240, China}
\affiliation[c]{Tsung-Dao Lee Institute, Shanghai Jiao Tong University, Shanghai 200240, China}
\affiliation[d]{Shanghai Jiao Tong University Sichuan Research Institute, Chengdu 610000, China
}
\affiliation[e]{Department of Energy and Power Engineering, School of Mechanical and Automotive Engineering,\\Shanghai University of Engineering and Science, Shanghai 201620, China
}

\emailAdd{yju@sjtu.edu.cn}

\abstract{The PandaX-30T is a proposed next-generation experiment to study dark matter and neutrinos using a dual-phase time projection chamber with \textasciitilde30 tons of liquid xenon. An innovative xenon handling subsystem of the PandaX-30T, the First-X, is described in this paper. The First-X is developed to handle liquid xenon safely and efficiently, including liquefying and long-term storing xenon without losses or contamination, and transferring cryogenic liquid xenon between the storage module and the detector safely and effectively without venting out. The storage module of the First-X is five specially designed double-walled cylindrical vessels (Center Tanks) equipped with three heat exchangers each for pressure and temperature regulation. Each Center Tank is designed with a vacuum and multi-layer insulation and a maximum allowable working pressure of 7.1 MPa, allowing 6 tons of xenon to be stored at 165--178 K at 0.1--0.2 MPa in the liquid phase or up to 300 K and up to 6.95 MPa in the supercritical phase. High-pressure storage (\textgreater0.2 MPa) only occurs in case of long-term detector shutdown or lack of nitrogen, ensuring no-loss storage of 6 tons of xenon in the range 178--300 K. In this paper, the thermophysical performances of the First-X and innovative scenario to conduct non-vented cryogen transportation were experimentally conducted and reported using liquid argon. The non-vented cryogenic liquid filling and pump-assisted cryogenic liquid recovery have been conducted with liquid argon at a mass flow rate of 1390 kg/h, corresponding to a xenon mass flow rate of 2140 kg/h. Compared with the PandaX-4T, the transportation of xenon between the detector and the storage module is conducted in the liquid phase rather than in the gaseous phase, and the filling rate (fill the detector) and the recovery rate (recover xenon from the detector) are increased by approximately 50 times and 30 times, respectively.
}

\begin{document}
\maketitle
\section{Introduction}\label{sec:intro}

Abundant astronomical evidence indicates the existence of dark matter, a non-relativistic and non-baryonic matter different from ordinary matter while contributing to the gravitational effect of the universe \cite{bertone_2005_particle, roos_2010_dark}. Numerous experiments have been carried out with telescopes, using colliders, and in underground laboratories to search for dark matter. To date, the direct detection method at the underground laboratories using liquid xenon (LXe) dual-phase time projection chambers (TPCs), such as PandaX \cite{pandaxcollaboration_2015_lowmass, pandax-iicollaboration_2016_dark, pandax-iicollaboration_2016_darka, pandax-iicollaboration_2017_dark, zhao_2018_pandax, pandax-iicollaboration_2022_search, pandax-4tcollaboration_2021_dark, pandaxcollaboration_2022_search, pandaxcollaboration_2022_first}, XENON \cite{aprile_2022_search}, and LZ \cite{aalbers_2022_first} are at the forefront in the quest for dark matter in forms of Weakly Interacting Massive Particles (WIMPs). LXe TPCs have been scaled up for several generations in recent years, in which the mass of xenon has increased substantially from 10 kg to 10 tons. For example, the PandaX collaboration has run PandaX-I \cite{pandaxcollaboration_2015_lowmass}, PandaX-II \cite{pandax-iicollaboration_2016_dark, pandax-iicollaboration_2016_darka, pandax-iicollaboration_2017_dark, pandax-iicollaboration_2022_search}, and PandaX-4T \cite{pandax-4tcollaboration_2021_dark, pandaxcollaboration_2022_search, pandaxcollaboration_2022_first} in China Jinping Underground Laboratory (CJPL) \cite{cheng_2017_china} with the mass of xenon of \textasciitilde120 kg, \textasciitilde580 kg, and \textasciitilde6 tons, respectively. Now, a 30--50 tons LXe TPC is proposed as the next generation of the dark matter and neutrinos experiment by PandaX-30T \cite{zhao_2018_pandax, cheng_2017_china} and Darwin \cite{aalbers_2016_darwin, baudis_2021_design}. Therefore, innovative changes are required to the xenon handling system so that the substantial increase in xenon mass can be addressed.

The major function of the xenon handling system is to manage xenon, between the external storage module and the detector, safely and effectively. Traditionally, the gaseous xenon (GXe) was transported from the gas storage module to the detector and subsequently liquified inside the detector by the cooling system. At the end of the experiment, the LXe in the detector was evaporated and then recovered to the storage module. However, the gas phase storage, filling, and recovery process is ineffective and unsuitable for handling significant quantities of xenon. As for the xenon storage, in PandaX \cite{gong_2013_cryogenic, zhao_2021_cryogenicsa}, LZ \cite{akerib_2020_luxzeplin} and XENON \cite{aprile_2012_xenon100}, xenon was stored in gas cylinders in the gaseous or supercritical phase at room temperature and high pressure, and the volume of each cylinder was only 40--50 L. PandaX-4T utilized as many as 128 gas cylinders to store \textasciitilde6 tons of xenon in the space-constrained underground laboratory \cite{zhao_2021_cryogenicsa}. Hence, as the mass of xenon increases to 30 tons, hundreds of cylinders and complex pipelines make the existing gas storage unsuitable for PandaX-30T. For the filling process, PandaX-4T, XENON1T, and LZ filled the detectors with GXe at a mass flow rate of 700 kg/d \cite{zhao_2021_cryogenicsa}, 50 SLPM (\textasciitilde380 kg/d) \cite{aprile_2017_xenon1t} and 40 SLPM (\textasciitilde300 kg/d) \cite{mount_2017_luxzeplin}, respectively, which was limited by the low density of the GXe, the size of the pipeline and valves, and the cooling power of the detector. In PandaX-4T and LZ, the mass recovery rate of xenon was only 200 SLPM (1533 kg/d) \cite{wang_2022_design}, and 300 SLPM (2300 kg/d) \cite{mount_2017_luxzeplin}, respectively. Furthermore, a heating system was needed to evaporate the LXe in the detector, and the balanced operation between the heating system and compressor became crucial for avoiding the overpressure of the detector or the freezing of xenon in the detector.

The storage, filling, and recovery can be performed in the liquid phase to make the system more compact and improve transportation efficiency. Xenon has a relatively high boiling point of 165 K, and it can be stored in the liquid phase at atmospheric pressure with proper thermal insulation techniques. In addition, with liquid storage, the xenon can be transported between the detector and the storage module in the liquid phase rather than in the gaseous phase. Since the density of liquid xenon (2854 kg/m$^3$ at 0.2 MPa, saturated) is hundreds of times that of the gaseous xenon (16 kg/m$^3$ at 0.3 MPa, 300 K), the mass flow rate can be increased significantly. XENONnT and XEMIS2 reported a fast liquid recovery at the mass flow rate of 1000 kg/h \cite{shingo_2018_xenonnt} by freezing the xenon in the ReStoX2 and 770 kg/h \cite{l.virone_2018_gravity} by gravity, respectively.

A novel cryogenic LXe handling system for the filling, recovery, and storage of xenon (the First-X) is designed for the PandaX-30T to increase transportation efficiency and system safety regarding the condition of CPJL. The major component of the First-X system is five specially designed double-walled cylindrical tanks (Center Tanks) equipped with three heat exchangers each for temperature and pressure regulation. Each Center Tank has an internal volume of 6.27 m$^3$, and is designed to store 6 tons of xenon. The Center Tank is designed with a vacuum and multi-layer insulation and a maximum allowable working pressure (MAWP) of 7.1 MPa, allowing 6 tons xenon to be stored in the liquid phase at 165--178 K at 0.1--0.2 MPa or up to 300 K and up to 6.95 MPa in the supercritical phase. In general, xenon is stored at 165--178 K at 0.1--0.2 MPa whilst the heat loss is compensated by the cooling modules. High-pressure (\textgreater0.2 MPa) storage only occurs in case of long-term detector shutdown or lack of LN2 in an emergency. In this case, LXe gradually evaporates and finally reaches ambient temperature due to inevitable heat losses. The ultimate Center Tank pressure with 6 tons of xenon stored at 300 K is 6.95 MPa, which is lower than the MAWP, 7.1 MPa, ensuring non-loss storage in the range 178--300 K. With the design of the First-X, the filling and the recovery process can be carried out in the liquid phase rather than in the gaseous phase as used in the current PandaX-4T, and the filling (filling the detector) and the recovery (recovery xenon from detector) rate are approximately 50 times and 30 times greater than that of the PandaX-4T, respectively, which greatly improves the efficiency and safety of the PandaX-30T experiment.

The structure of the present paper is organized as follows: after a brief introduction to the previous xenon handling system in Section \ref{sec:intro}, the detailed design and construction of the First-X is introduced in Section \ref{sec:design}. Experimental results and discussions are presented in Section \ref{sec:results}, followed by the conclusion, in Section \ref{sec:conclusion}.

\section{Design and construction of the First-X}\label{sec:design}

\subsection{System overview} \label{sec:overview}

As outlined in Section \ref{sec:intro}, the primary function of the First-X is to liquefy xenon inside the Center Tanks, to store xenon long-term outside the detector, to quickly fill the detector with xenon, and to recover xenon from the detector to the Center Tanks safely and efficiently. As part of the low-background experimental facilities at the underground laboratory, the following requirements are proposed for the First-X:

\begin{itemize}
\item
  Store 30 tons xenon in five Center Tanks in the liquid phase at cryogenic temperatures and in the gaseous phase at ambient temperature. The heat loss during the cryogenic storage should be less than 150 W. Therefore, \textasciitilde300 L nitrogen per day is required to ensure a stable storage of 30 tons of xenon (assuming the outlet temperature of the LN2 cooling module is 150 K).
\item
  Fill five Center Tanks successively with gaseous xenon from the GXe bottles at a mass flow rate of 500 SLPM (\textasciitilde160 kg/h), and simultaneously liquefy xenon at 178 K, 0.2 MPa (In this paper, the pressure refers to absolute pressure) with a cooling power of approximately 5000 W.
\item
  Fill the detector with LXe, without venting, effectively at a mass flow rate of more than 1500 kg/h whilst preventing detector over pressurization.
\item
  Recover LXe from the detector and transfer to Center Tanks at a mass flow rate of more than 1500 kg/h with a lifting head of approximately 12 meters without venting, after detector shutdown or in an emergency.
\item 
Recover GXe in order to release the pressure of the detector when the detector pressure rises abnormally for a short period of time.
\item 
Reduce contamination during storage. All system components in contact with xenon should be cleaned with alcohol to remove surface stains. The helium leak rate of all the seals should be lower than 1\(\times\)10\(^{- 10}\) Pa\(\cdot\)m\(^{3}\)/s to reduce the air leak contamination, and the inner surface roughness of the pipeline and the storage module should be less than 0.4 \(\mu\)m to reduce the outgassing rate, thereby reducing outgassing contamination.
\end{itemize}

\begin{figure}
  \centering
  \includegraphics[width=1.0\textwidth]{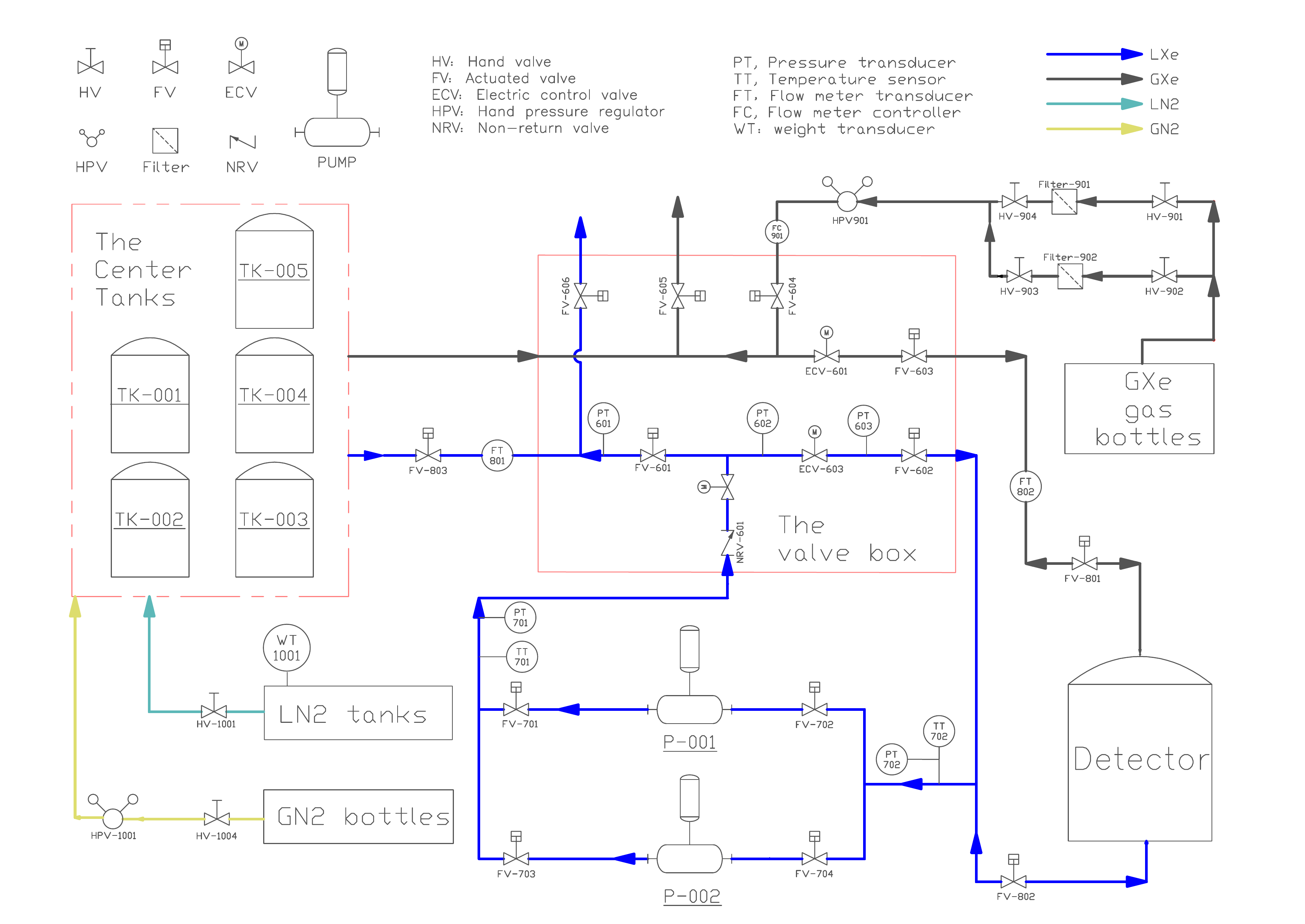}
  \caption{Simplified process and instrumentation diagram of the First-X.}
  \label{fig:flow_chart}
  \end{figure}

\begin{landscape}
\begin{figure}
  \centering
  \includegraphics[width=1.5\textwidth]{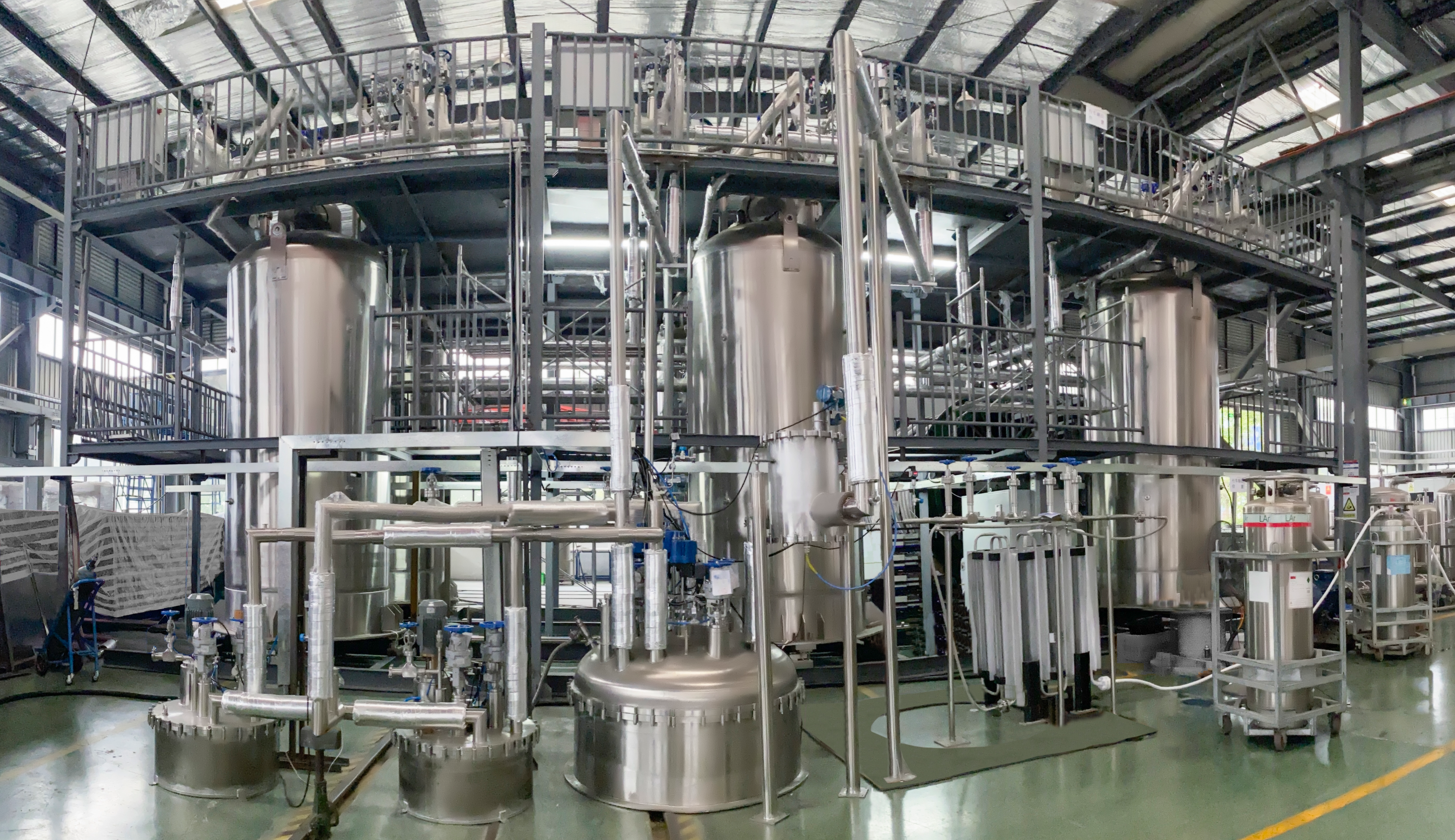}
  \caption{A photograph of the First-X.}
  \label{fig:pho}
  \end{figure}
\end{landscape}

The First-X, represented schematically in figure \ref{fig:flow_chart} and shown in figure \ref{fig:pho}, consists of xenon gas bottles from the supplier, a valve box for distributing the xenon, five Center Tanks for xenon storage, two cryogenic pumps (one for backup) to recover LXe, and LN2 and GN2 storage modules for pressure and temperature regulation. The gas bottles are commercial high-pressure cylinders storing the xenon in the gaseous or supercritical phase. At the beginning of the experiments, the GXe is transported from the gas bottles to the Center Tanks and simultaneously liquified inside the Center Tanks. Five Center Tanks are designed to store 30 tons of xenon. Detailed parameters of the Center Tanks are described in Section \ref{sec:center_tank} and Section \ref{sec:heat_exchanger}. The valve box, the LXe pipelines (blue lines in the figure \ref{fig:flow_chart}), and the GXe pipelines (black lines in the figure \ref{fig:flow_chart}) are also vacuum and multi-layer insulated to reduce heat losses. If the detector is ready for filling, LXe is driven from the Center Tanks to the detector by its pressure difference without venting. During the non-vented filling (NVF), the Center Tank is pressurized by injecting GXe from the xenon gas bottles whilst the detector is properly controlled without over pressurization. In the event of detector shutdown or an emergency, LXe is recovered from the detector and transferred to the Center Tanks by the cryogenic pumps. The filling and recovery are detailed in Section \ref{sec:nvf_nvr}. Industrial programmable logic system controllers (PLCs) are designed and constructed to handle the read-out and the control strategies. Figure \ref{fig:pho} is a photograph of the First-X with five Center Tanks (Two Center Tanks are hidden due to the viewing angle), a valve box, and two cryogenic pumps. The facility covers an area of 120 m\(^{2}\).

\subsection{The storage module---Center Tanks} \label{sec:center_tank}

The First-X consists of five Center Tanks designed to store 6 tons of xenon each. The Center Tank is made of stainless steel with an inner volume of 6.27 m\(^{3}\). The Center Tank is a cylindrical double-walled cryogenic vessel with a vacuum and multi-layer insulation for cryogenic liquid storage, as shown in the cross-section view of the Center Tank in figure \ref{fig:cross_section}(a). Additionally, the Center Tank has a MAWP of 7.1 MPa to allow high-pressure storage.

In general, LXe is stored at 0.1--0.2 MPa at 165--178 K by operating the LN2 cooling circuit or condenser, which are described in Section \ref{sec:heat_exchanger}. The liquid level of the 6 tons of xenon inside the Center Tank is about 1.31 m, and the liquid volume is approximately 35\% of the total inner volume. Therefore, the Center Tanks can store more LXe by increasing the filling ratio when the xenon is stored at cryogenic temperatures. During the stable liquid storage process, the Center Tanks can be connected to the purification modules, such as the distillation tower, to remove the associated contamination.

During the detector shutdown or long-term lack of nitrogen, xenon can be stored in the Center Tanks without cooling. In this period, LXe gradually evaporates and finally reaches ambient temperature due to the inevitable heat losses. Even if 6 tons of LXe (1.31 m height) evaporates and reaches 300 K, the maximum pressure of the Center Tank is 6.95 MPa, which is lower than the MAWP of 7.1 MPa. The experiment hall is located under the Jinping Mountain, with 2,400 m of rock overburden, and as a result, the temperature of the experiment hall has been maintained at a relatively constant temperature of approximately 20 °C throughout years of operation. The laboratory has strict fire prevention measures, and large-scale fires will not occur, which significantly reduces the likelihood of a rise in the ambient temperature. Each Center Tank is equipped with two sets of safety components, which are backups of each other, ensuring that one can always be used. Each set of safety components consists of a safety valve and a burst disc connected in series. In order to reduce the leakage of the safety valve, the outlet of the rupture disc is connected to the inlet of the safety valve. The burst disc of the Center Tank will rupture when the Center Tank pressure reaches 7.1 MPa (6 tons xenon at 302 K). As the ambient temperature continues to increase, the safety valve will open when the Center Tank pressure reaches 7.35 MPa to remove the pressure and ensure the pressure of the Center Tank is always lower than 7.35 MPa.

With a design pressure of 7.8 MPa (1.1 times greater than the MAWP), the Center Tank inner vessel wall thickness is 40 mm and the weight of the inner vessel is up to 6 tons. The heavy inner vessel results in the cool-down process taking longer. However, the heavy inner vessel can act as a substantial thermal buffer, providing cooling power to the system after it is completely cooled down. Specifically, the inner vessel can release 2280 kJ of cooling capacity by rising 1 K while the specific heat capacity of stainless steel is 380 J/kg/K \cite{NIST304} at 165 K. The Center Tanks are sealed by the copper gasket in a specially designed CF flange. Through the large top neck with an inner diameter of 448 mm, the heat exchangers can be inserted into the Center Tanks, and the inner surface of the inner vessel can be polished and cleaned. The inner surface of the vessel is mechanically polished to a roughness of less than 0.4 \(\mu\)m. All the pipes connected to the top flange are vacuum jacketed to reduce heat losses and avoid the low temperature of the top flange. Two radar liquid level meters (DOWESTON, range: 0--5 m, accuracy: 10 mm) are mounted inside each Center Tank to measure the liquid level. Four weight sensors (range: 0--30 tons, accuracy: 15 kg) are installed under each Center Tank to measure the mass of the liquid and Center Tank. Several PT100 temperature sensors (accuracy: 0.5 K) are mounted at the outer surface of the inner vessel of each Center Tanks ($h =0$ m, 1.45 m and 3.66 m) and inside the Center Tanks ($h =0$ m, 1.1 m, 1.25 m, 1.35 m, 2 m, 2.8 m, 3.6 m) to monitor the inner vessel temperatures and LXe/GXe temperatures. A pressure transducer and a pressure gauge are installed on the pipeline connected to the gas phase of each Center Tank to monitor the pressure (range: 0--10 MPa, accuracy: 0.01 MPa). The essential parameters of the Center Tank are listed in Table \ref{tab:spe}.

\begin{table}
  \centering
  \caption{Specifications of the Center Tank.}
  \setlength{\tabcolsep}{10mm}{
  \begin{tabular}{lr}
  \toprule
  Dimensions & 1800 (diameter) $\times$4700 (height) mm \\
  \midrule
  Inner volume & 6.27 m$^3$ \\
  Design pressure & 7.8 MPa \\
  Hydrostatic test pressure & 9.8 MPa \\
  Design temperature & 77$\sim$313 K \\
  Operation temperature & 178 K \\
  Roughness of the inner surface &\textless0.4 $\mu$m \\
  Inner diameter of the neck & 448 mm \\
  Thickness of the inner vessel walls & 40 mm \\
  Mass of the inner vessel & $\sim$6 tons \\
  \bottomrule 
  \end{tabular}
  }
  \label{tab:spe}
  \end{table}

\subsection{Heat exchangers} \label{sec:heat_exchanger}

\begin{figure}[htbp]
  \centering
  \subfloat[]{
  \begin{minipage}[c]{0.49\textwidth}
  \centering
  \centerline{\includegraphics[width=10cm,height=8cm,keepaspectratio=True]{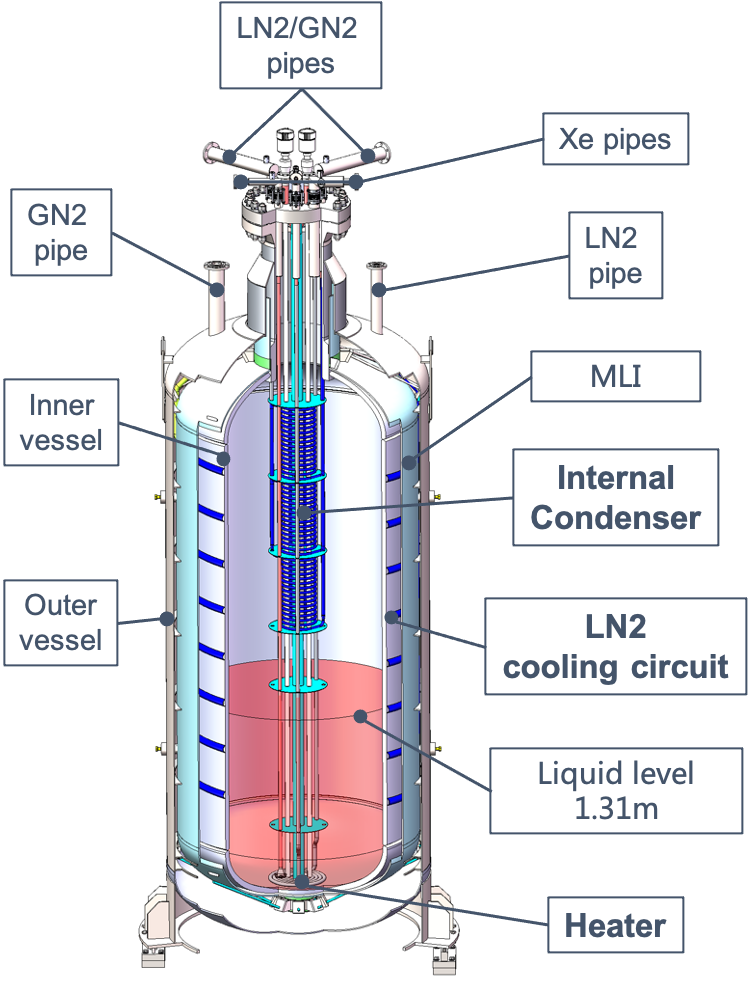}}
  \end{minipage}
  }
  \centering
  \subfloat[]{
  \begin{minipage}[c]{0.49\textwidth}
  \centering
  \centerline{\includegraphics[width=10cm,height=8cm,keepaspectratio=True]{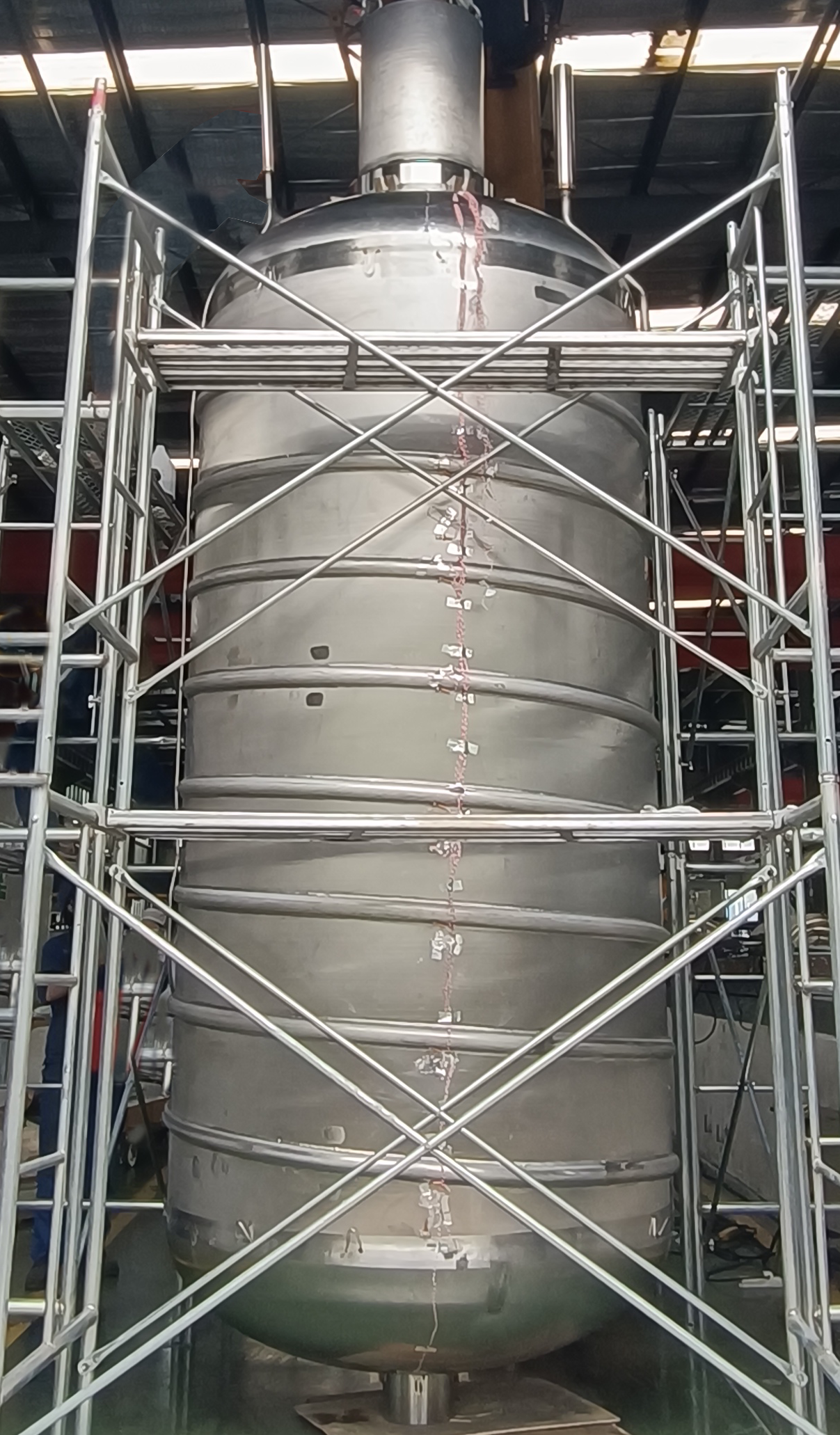}}
  \end{minipage}
  }
  \\ 
  \centering
  \subfloat[]{
  \begin{minipage}[c]{0.98\textwidth}
  \centering
  \centerline{\includegraphics[width=12cm,height=4cm,keepaspectratio=True]{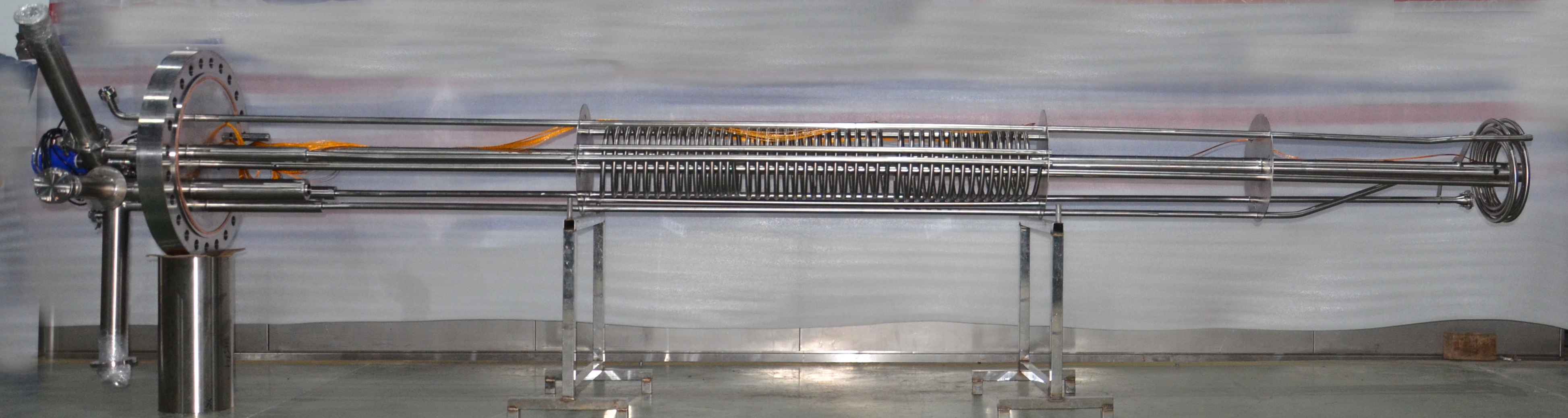}}
  \end{minipage}
  }
  \caption{(a) The cross-section view of the Center Tank. The Center Tank is double-walled with a vacuum and multi-layer insulation. The internal condenser, LN2 cooling circuit and Heater are shown as bold; (b) The internal vessel of the Center Tank, the irregular vertical and horizontal steel in front are temporary support structures. The semi-circular profile tube welded on the outside of the cylindrical part of the inner vessel is the LN2 cooling circuit; (c) The condenser and the heater in a horizontal position. The condenser is the stainless-steel coil at the center of the picture and the heater is the stainless-steel coil at the right side of the picture.}
  \label{fig:cross_section}
  \end{figure}

Each Center Tank is equipped with three main temperature control elements named Internal condenser, LN2 cooling circuit, and Heater, shown as bold in figure \ref{fig:cross_section}(a). The cooling medium is provided by one 6 m$^{3}$ and two 0.5 m$^{3}$ cryogenic LN2 vessels, whist the heating medium is provided by 20 GN2 gas bottles.

The LN2 cooling circuit is made of a single 40 m semi-circular profile tube which is welded on the outer surface of the inner vessel in an upward spiral way at the height from 0.4 m to 3.1 m, as shown in figure \ref{fig:cross_section}(b). The cooling coil has an inner diameter of 45 mm and a wall thickness of 3 mm.

The condenser and heater are shown in figure \ref{fig:cross_section}(c). Both the condenser and the heater are made of stainless steel tube with a diameter of 18 mm and wall thickness of 1.5 mm. The condenser and the heater are mechanically polished and alcohol cleaned to allow xenon storage without sacrificing its purity.

The condenser is located inside the Center Tank and is directly in contact with the GXe. Therefore, the Center Tank pressure can be reduced by liquefying the xenon when it is higher than expected. The condenser is made of three stainless tube coils winded in a circular helix shape. The pitch and the radius of the circular helix are 33 mm and 125 mm, respectively. The total length of the tube is 35 m.

The GN2-based heater is installed at the bottom of the Center Tank and in contact with LXe directly to melt any xenon ice (xenon ice may be present due to the operation of the cooling circuit or condenser) or evaporate LXe and fill the detector with GXe. The heater is made of two stainless tube coils winded in an Archimedean spiral shape. The distance between the loop is approximately 20 mm. The total length of the tube is approximately 6 m.

The heating power of the heater is designed as 2000 W with GN2 mass flow rate at 69 kg/h, and the inlet and outlet temperature of the GN2 is designed at 300 K and 200 K, respectively. The mass flow rate of the heater is adjusted by the GN2 mass controller installed upstream. It takes 110 minutes to deplete 20 gas bottles. The Center Tank pressure can be increased at a rate of 17 kPa/h using the heater when 6000 kg of xenon is stored at atmospheric pressure.

The cooling circuit and the condenser depressurize the Center Tank differently. For the LN2 cooling circuit, LN2 is injected into the circuit at the height of 0.4 m where it flows in direct contact with the Center Tank inner vessel external wall. Therefore, the temperatures of the inner vessel are the first to drop. Next, the Center Tank pressure drops due to the heat transfer between the LXe/GXe and the inner vessel. For the condenser, the GXe contacting the condenser is cooled and liquified at the beginning, resulting in a drop in the Center Tank pressure. Then, the temperatures of the inner vessel are subsequently reduced due to the heat transfer between the GXe and the inner vessel. Therefore, the LN2 cooling circuit is designed for long-term storage, while the condenser is designed to quickly reduce the Center Tank pressure.

The cooling power of the LN2 cooling circuit and the condenser can be regulated by adjusting the mass flow rate of the LN2. The inlet temperature of LN2 is approximately 77 K, and the exhaust GN2 temperature varies according to the inner vessel temperatures and LN2 mass flow rate. For each cooling module, the exhaust GN2 mass flow rate is measured by a Venturi GN2 mass flow meter installed at the outlet and controlled by adjusting the upstream LN2 control valve. A temperature sensor is mounted at the inlet and outlet of each cooling module to calculate the cooling power.

In order to avoid the freezing of the xenon, both the LN2 cooling circuit and the condenser are designed to intermittently turn on and turn off by the PLCs. A temperature sensor mounted at the outer surface of the inner vessel at a height of 1.45 m, $T_{h=1.45\, \rm{m}}$, and a pressure transducer connected to the gaseous phase of the Center Tank, $P_{\rm{Center \, Tank}}$, are used as the trigger for the LN2 cooling circuit and the condenser, respectively. By setting the upper and lower trigger points, the inlet valve of the cooling module is turned on when $T_{h=1.45\, \rm{m}}$ or $P_{\rm{Center \, Tank}}$ is greater than the upper trigger point, and the inlet valve is turned off when $T_{h=1.45\, \rm{m}}$ or $P_{\rm{Center \, Tank}}$ is less than the lower trigger point. The Center Tank can be stabilized at any desired pressure \(P_{s}\) by setting the cooling circuit trigger temperatures to (\(T_{s} -\)1 K) and \(T_{s}\) (where \(T_{s}\) is the saturation temperature of \(P_{s}\)) or setting the condenser trigger pressures to (\(P_{s} -\)0.01 MPa) and \(P_{s}\). The trigger points can be set as needed. The stable storage can be achieved by operating the LN2 cooling circuit or the condenser. During the stable storage, the trigger temperatures of the cooling circuit can be set to 165 K and 178 K, and the trigger pressures of the condenser can be set to 0.1 MPa and 0.2 MPa. Before the NVF, the cooling circuit trigger temperatures can be set to 164 K and 165 K to minimize the initial internal energy of the LXe.

\subsection{Non-vented filling and recovery} \label{sec:nvf_nvr}

\begin{figure}
  \centering
  \includegraphics[width=0.98\textwidth]{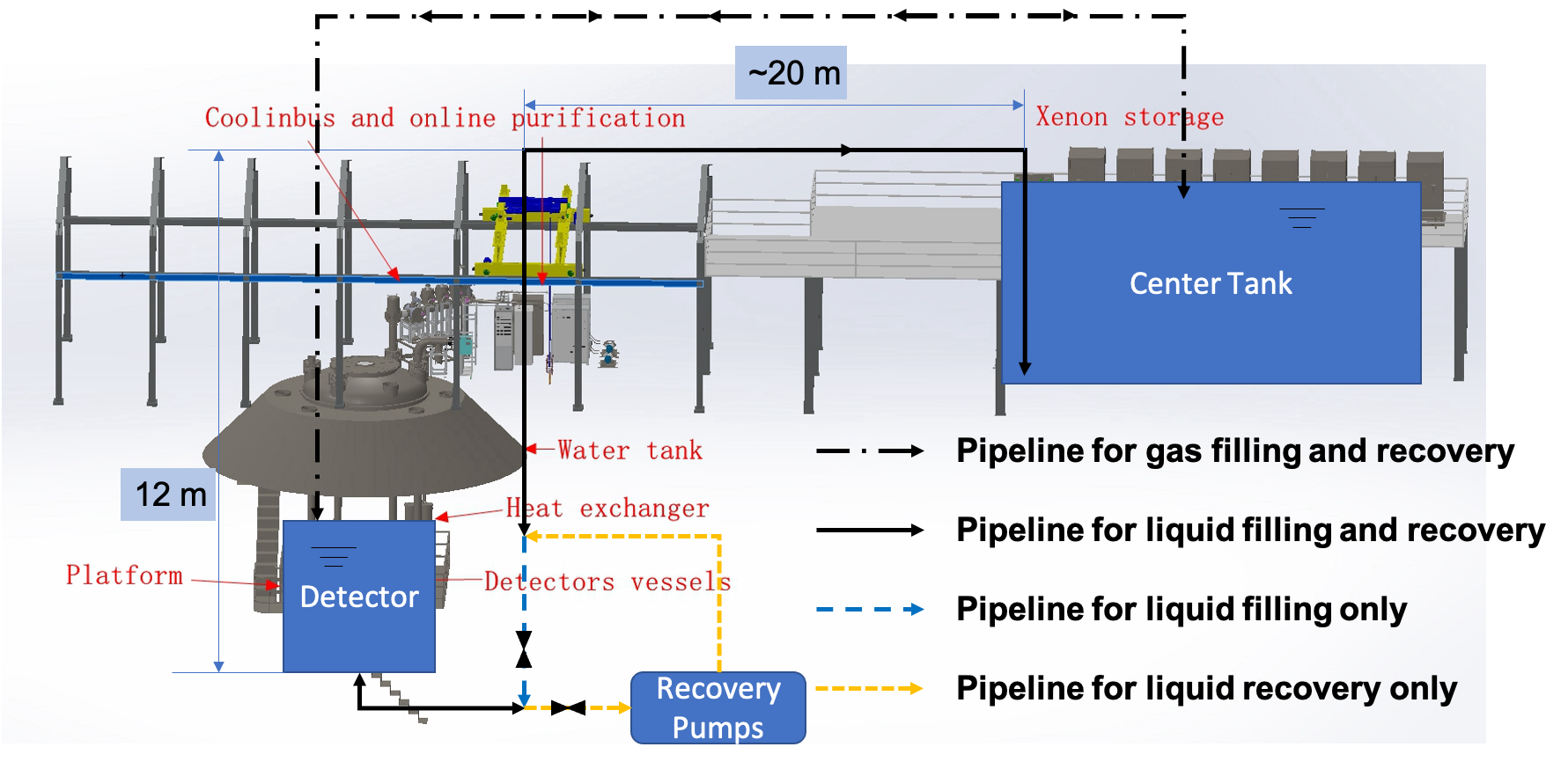}
  \caption{The relative position of the detector and First-X storage module for PandaX-30T (not to scale).}
  \label{fig:relative}
  \end{figure}

In PandaX-30T, the relative position of the detector and Center Tanks is the same as it was in PandaX-4T \cite{zhao_2021_cryogenicsa}. As shown in figure \ref{fig:relative}, the detector is located inside the water pit and is about 12 m below the liquid outlet port of the Center Tanks. The horizontal distance between the detector and Center Tanks is approximately 20 m. The ullage phase of the detector and Center Tanks are connected by a pipeline for gas filling and recovery (dash dot black line). Most of the pipelines for the liquid filling and recovery are shared (solid black line). The liquid pipeline is divided into two branches near the detector, one for filling (blue dash line) and one for recovery (yellow dash line). Since the Center Tank can withstand high pressures, the filling process is initiated by pressurizing the Center Tank, and the LXe flows into the detector driven by the pressure difference between the Center Tank and the detector. However, for the recovery process, this method cannot be used. In general, LXe is stored in the detector in saturated at 0.2 MPa, and the maximum working pressure is 0.35 MPa (on reaching this pressure the burst disc of the detector will rupture). As a result, the LXe cannot be lifted by 12 m (In this case, the detector pressure needs to be at least 0.33 MPa higher than the saturation pressure of LXe in the detector) and recovered to the Center Tank by pressurizing the detector, so the cryopump is designed to drive the recovery process. In PandaX-30T, two cryogenic pumps for the recovery are specially designed and manufactured by Barber Nichols \cite{bncom} (BNCP-32H-050, U.S.A.). The cryogenic pump is a centrifugal pump with a magnetic drive to isolate the pump and the LXe from the electric motor. Only the pump shaft with the impeller and driven section of the magnetic coupling are in contact with the xenon, resulting in no sacrificing of the purity of the xenon. The cryogenic pump is equipped with a motor and a variable frequency drive which allows adjustment of the pump speed to produce the desired head up to 0.57 MPa and the xenon mass flow rate up to 3900 kg/h within the available power range of the motor. With a maximum frequency of 60 Hz, the cryogenic pump can provide a head of 20 m with a LXe mass flow rate of 1500 kg/h. The cryogenic pump is used commercially globally with high reliability. In the PandaX-30T experiment, it takes approximately 20 hours to recover 30 tons of LXe at a mass flow rate of 1500kg/h. However, the system is equipped with a backup pump for extreme situations.

\subsubsection{Non-vented liquid filling (NVF)} \label{sec:nvf}

\begin{figure}
  \centering
  \includegraphics[width=0.9\textwidth]{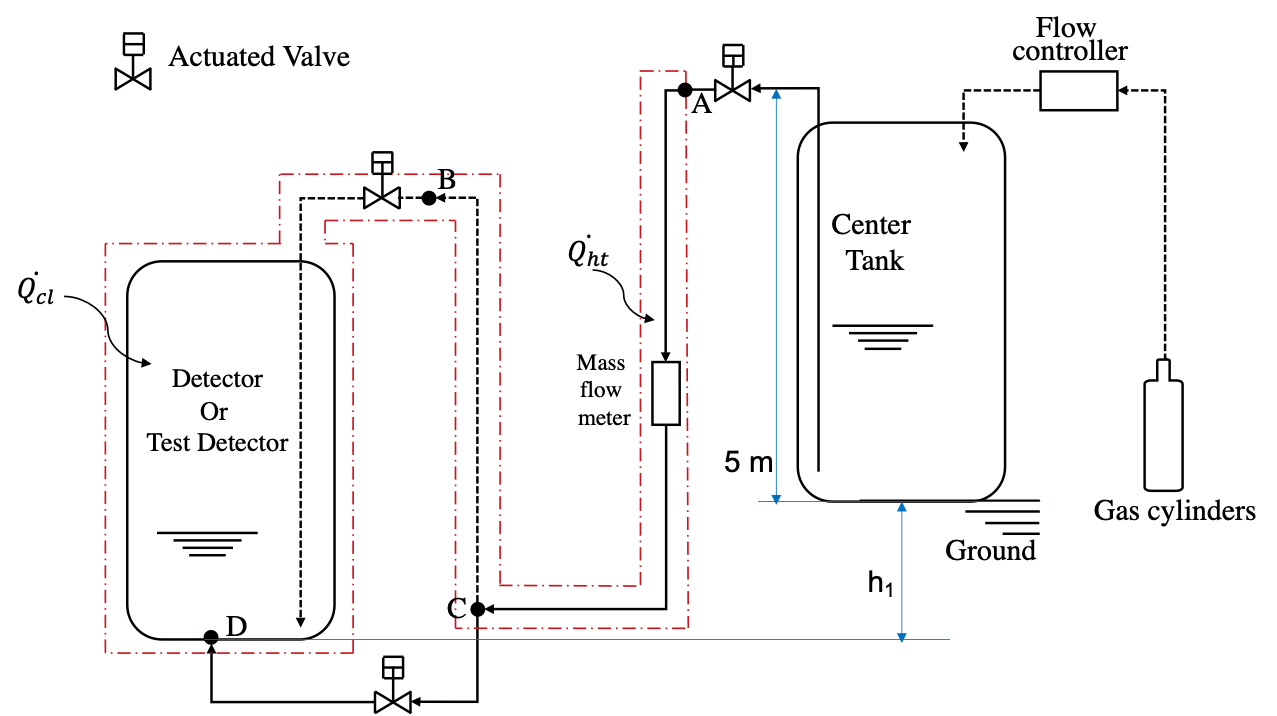}
  \caption{Simplified schematic for the filling process.}
  \label{fig:sche_filling}
  \end{figure}

As mentioned above, the liquid NVF process is designed for PandaX-30T in order to increase the mass flow rate. The fluid flow schematic during the filling is shown in figure \ref{fig:sche_filling}, which includes the gas bottle for pressurization, the Center Tank, the detector or Test Detector (in the preliminary experiments, one of the Center Tanks acted as the detector, named Test Detector), and the transfer line between the vessels. In PandaX-30T, the liquid inlet of the detector is located at the bottom and is 12 m below the outlet of the Center Tank point A (\(h_{1} = 7\) m). The transfer line is the solid line between the outlet of the Center Tank point A and point C, plus the solid line between the point C and the inlet of the detector point D. Since one of the Center Tanks acted as the Test Detector in the preliminary experiments, the liquid outlet was at the top of the Test Detector instead at the bottom. Additionally, the Center Tank and Test Detector were on the same level (\(h_{1} = 0\) m). Therefore, the transfer line was the solid line between the point A and point C, plus the dash line between point C and the Test Detector.

Two main requirements are needed in the NVF: establishing a pressure difference between the Center Tank and the detector to drive the fluid transportation, and avoiding the overpressure of the detector. The pressure of the Center Tank can be increased and controlled by injecting GXe (\textasciitilde300 K) from the gas cylinders into the ullage of the Center Tank. Through a PLC command, the mass controller installed upstream of the gas supply line can be turned on and off. Thereby, the Center Tank pressure can be controlled within its setting range to overcome the flow resistance between the Center Tank and the detector.

The MAWP of the detector is as low as 0.35 MPa, and the overpressure must be avoided as it can lead to fatal damage to the photomultiplier tubes and the sensors inside the detector. Since the heat loss of the pipeline and the detector \({\dot{Q}}_{ht}\) are almost constant, there are three methods to lower the pressure of the detector: (1) reduce the initial xenon temperature in the Center Tank while avoiding freezing the xenon before NVF; (2) precool the detector before NVF; (3) increase the cooling power of the detector \({\dot{Q}}_{cl}\) during the NVF while avoiding freezing the xenon.

\subsubsection{Non-vented liquid recovery (NVR)} \label{sec:nvr}

\begin{figure}
  \centering
  \includegraphics[width=0.9\textwidth]{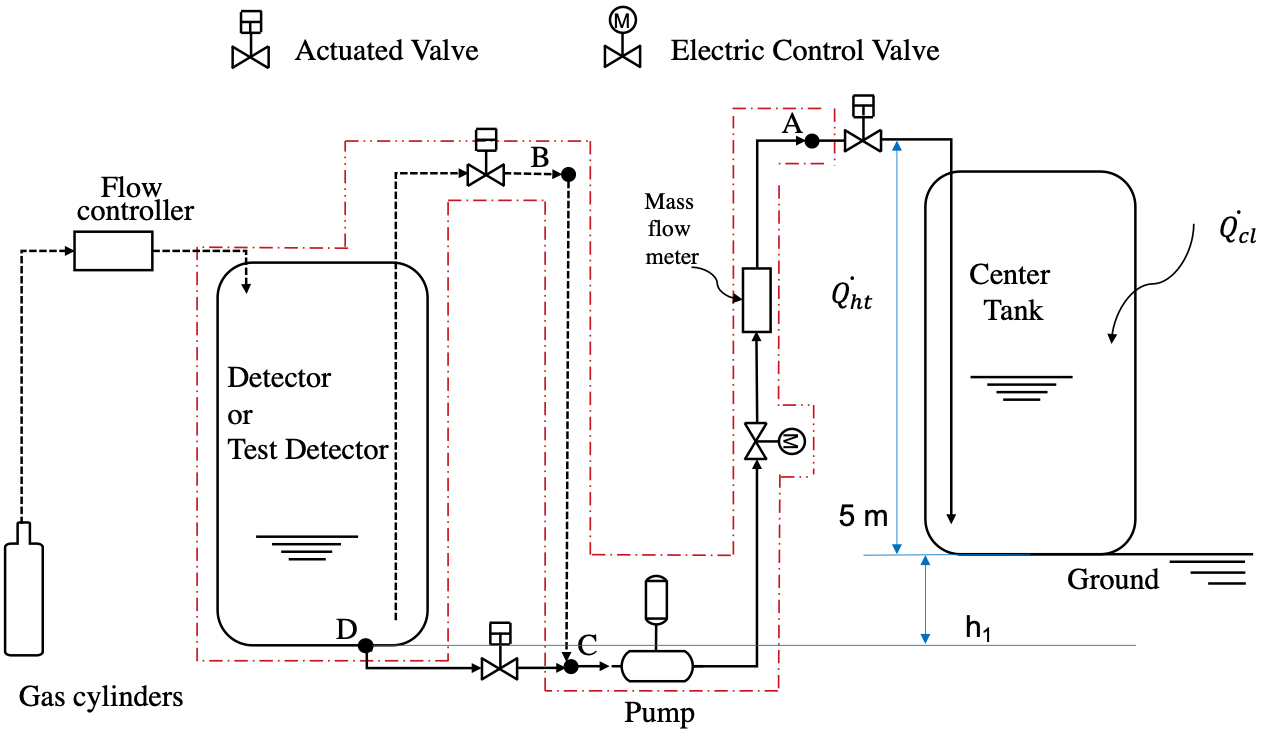}
  \caption{Simplified schematic for the recovery.}
  \label{fig:sche_recovery}
  \end{figure}

The fluid flow schematic during non-vented recovery (NVR) is shown in figure \ref{fig:sche_recovery} which includes the gas cylinders, the detector or Test Detector, the pumps, the Center Tank, and the transfer lines (black solid lines) between the vessels. According to the geometry and infrastructure in PandaX-30T, the outlet of the detector (point D) is expected to be 12 m lower than the liquid outlet of Center Tank point A (\(h_{1} = 7\) m). Since the MAWP of the detector is 0.35 MPa, it is impossible to increase the pressure of the detector and then push the LXe from the detector into the Center Tank. Therefore, two cryogenic LXe pumps (one for backup) are introduced to achieve recovery.

In PandaX-30T, the liquid outlet of the detector is located at the bottom of the detector. As the recovery begins, LXe flows from the outlet of the detector point D through the black solid line to the cryogenic pump by gravity, and is then pumped to the Center Tank. During the NVR, the pressure of the detector is stabilized by the evaporated gas caused by the detector heat losses. The transfer line is the solid line between point D and point A. In the preliminary experiments with liquid argon (LAr) instead of xenon, the Center Tank and detector were on the same level (\(h_{1} = 0\) m). The liquid outlet of the Test Detector was at the top rather than at the bottom. In order to push the LAr inside the Test Detector to the cryogenic pumps, the pressure of the Test Detector was increased by injecting the gaseous argon (GAr) from the gas bottles into the ullage, as displayed in the left side of figure \ref{fig:sche_recovery} in black dash lines. A control valve was installed downstream of the pump to simulate the pressure drop caused by height difference.

During the recovery, the temperature of the LXe in the detector is constant and cannot be adjusted quickly. Hence, two strategies can be used to reduce the pressure of the Center Tank: (1) precool the inner vessel of the Center Tank before the NVR; (2) increase the cooling power of the Center Tank during the NVR.

\section{Experimental results and discussion} \label{sec:results}

\subsection{Overview of the preliminary test} \label{sec:exp_overview}

After the First-X was constructed, experiments were carried out using less costly argon rather than expensive xenon. Argon has a higher boiling point (87.4 K) than nitrogen, so the pressure of the Center Tank can be reduced by liquefying argon with the LN2 cooling modules. Argon also has a higher freezing point (83.9 K) than nitrogen. Therefore, strategies to avoid the freezing of LXe can be applied to LAr. Additionally, argon is chemically inert and can be easily pumped out after the experiments, making it suitable for preliminary test.

In the preliminary test, one Center Tank acted as the liquid storage module, and the other Center Tank acted as the Test Detector. As a result, the Center Tank and Test Detector were on the same level, which was different from the actual situation of the PandaX-30T. However, in this paper, we focus on the thermophysical performance of the Center Tank and Test Detector, whereas the hydraulic performance of the system related to the relative position will be investigated in the future.

The experiments were conducted in two stages. In the first stage, \textasciitilde3000 kg of argon was filled into the Center Tank. The heat losses of the Center Tank were tested, followed by pressure regulation experiments using three heat exchangers. The results are discussed in Section \ref{sec:exp_storage} and Section \ref{sec:exp_pre_reg}. In the second stage, only \textasciitilde2600 kg of cryogenic LAr was filled in the Center Tank to conduct the NVF and NVR experiments. The results of the two-stages experiments are presented in detail in Section \ref{sec:exp_nvf_nvr}.

\subsection{Storage} \label{sec:exp_storage}

\subsubsection{Cool-down the Center Tank} \label{sec:cool-down}

The experiments started with the cool-down of the Center Tank using the LN2 cooling circuit. The inlet of the LN2 circuit is located at a height of 400 mm. Therefore, the inner vessel close to the LN2 inlet was cooled down more rapidly than the bottom and the upper part of the inner vessel. Therefore, in order to cool down the bottom part of the inner vessel quickly, the GAr was filled into the Center Tank. As the GAr was cooled and liquefied by the cooler part of the inner vessel and then descended to the bottom due to gravity, the bottom part of the inner vessel was continuously cooled. The control strategy described in Section \ref{sec:heat_exchanger} was adopted with the trigger temperatures set as 86 K and 87 K to avoid the argon freezing.

\begin{figure}[H]
  \centering
  \includegraphics[width=0.6\textwidth]{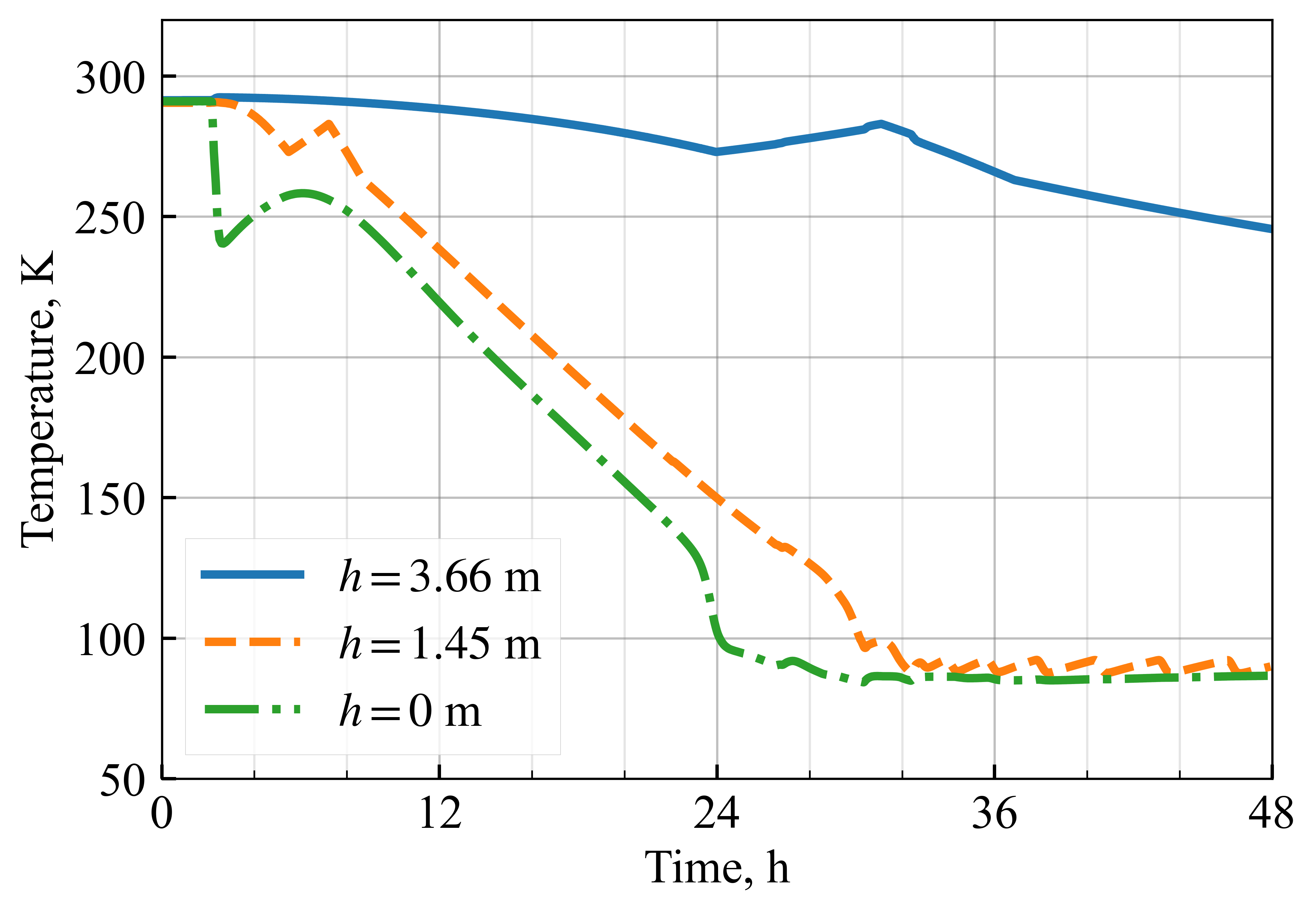}
  \caption{Inner vessel temperatures at $h =0$ m, 1.45 m, and
  3.66 m during the cool-down process.}
  \label{fig:precool}
  \end{figure}

The inner vessel temperatures at $h =0$ m, 1.45 m, and 3.66 m are displayed in figure \ref{fig:precool}. At the beginning of the cool-down, the cold GAr filled the Center Tank through the inlet of the liquid pipe. As a result, the temperature at the $h =0$ m decreased at the beginning. As the cooling process proceeded, the heat transfer between the LN2 and the inner vessel at the front circuit changed from film boiling to nuclear boiling, causing the temperatures at $h = 1.45$ m and 3.66 m to increase slightly.

As shown in figure \ref{fig:precool}, the Center Tank cooled down in approximately 32 hours until the inner vessel temperature at $h = 1.45$ m and $h =0$ m dropped to 87 K. Despite the large heat capacity of the inner vessel, the average temperature of the inner vessel dropped at an average of 3.44 K/h. Additionally, at the end of the cool-down, the temperature difference between the temperature at $h =3.66$ m and $h =0$ m was as high as 159 K. If the inner vessel temperature at $h =3.66$ m was lower, the heat loss of the top neck due to the conduction would be significantly increased. Since the ultimate liquid level is about 1.31 m, it is unnecessary to precool entire inner vessel.

\begin{figure}[H]
  \centering
  \includegraphics[width=0.6\textwidth]{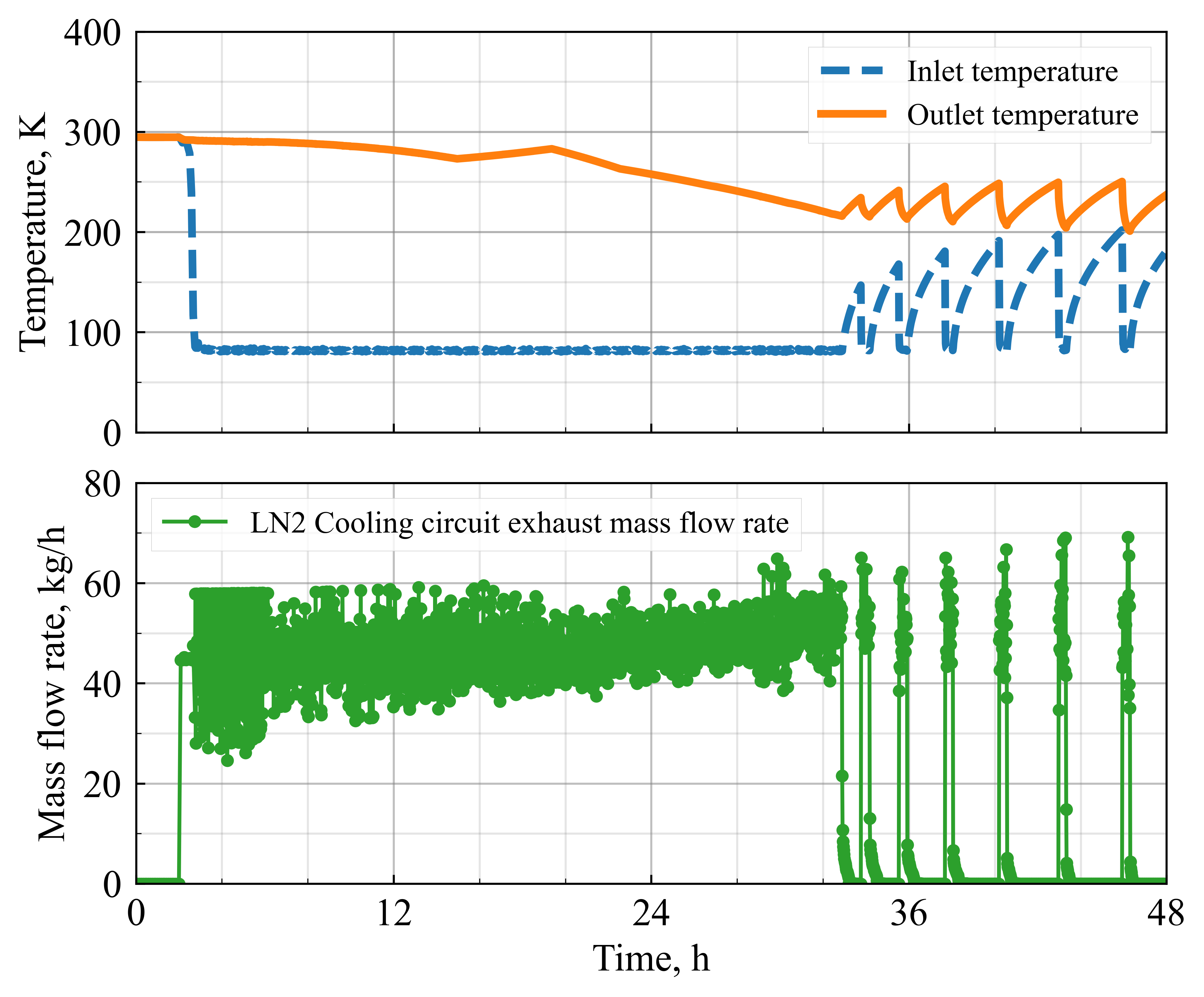}
  \caption{The inlet and outlet temperatures of the LN2 cooling
  circuit, and the exhaust GN2 mass flow rate at the outlet of the cooling
  circuit during the cool-down process.}
  \label{fig:precool_nitro}
  \end{figure}

The inlet and outlet temperatures as well as the exhaust GN2 mass flow rate of the cooling circuit are displayed in figure \ref{fig:precool_nitro}, which shows that the outlet temperature was greater than 200 K, indicating a relatively high efficiency of the LN2 cooling circuit. Also, the GN2 mass flow rate fluctuated within a large range due to the boiling transfer between the LN2 and the Center Tank inner vessel. The maximum and average cooling power of the LN2 cooling circuit were as great as 6700 W and 3600 W, respectively.

\subsubsection{Liquefaction of gaseous argon} \label{sec:liq}

\begin{figure}
  \centering
  \includegraphics[width=0.6\textwidth]{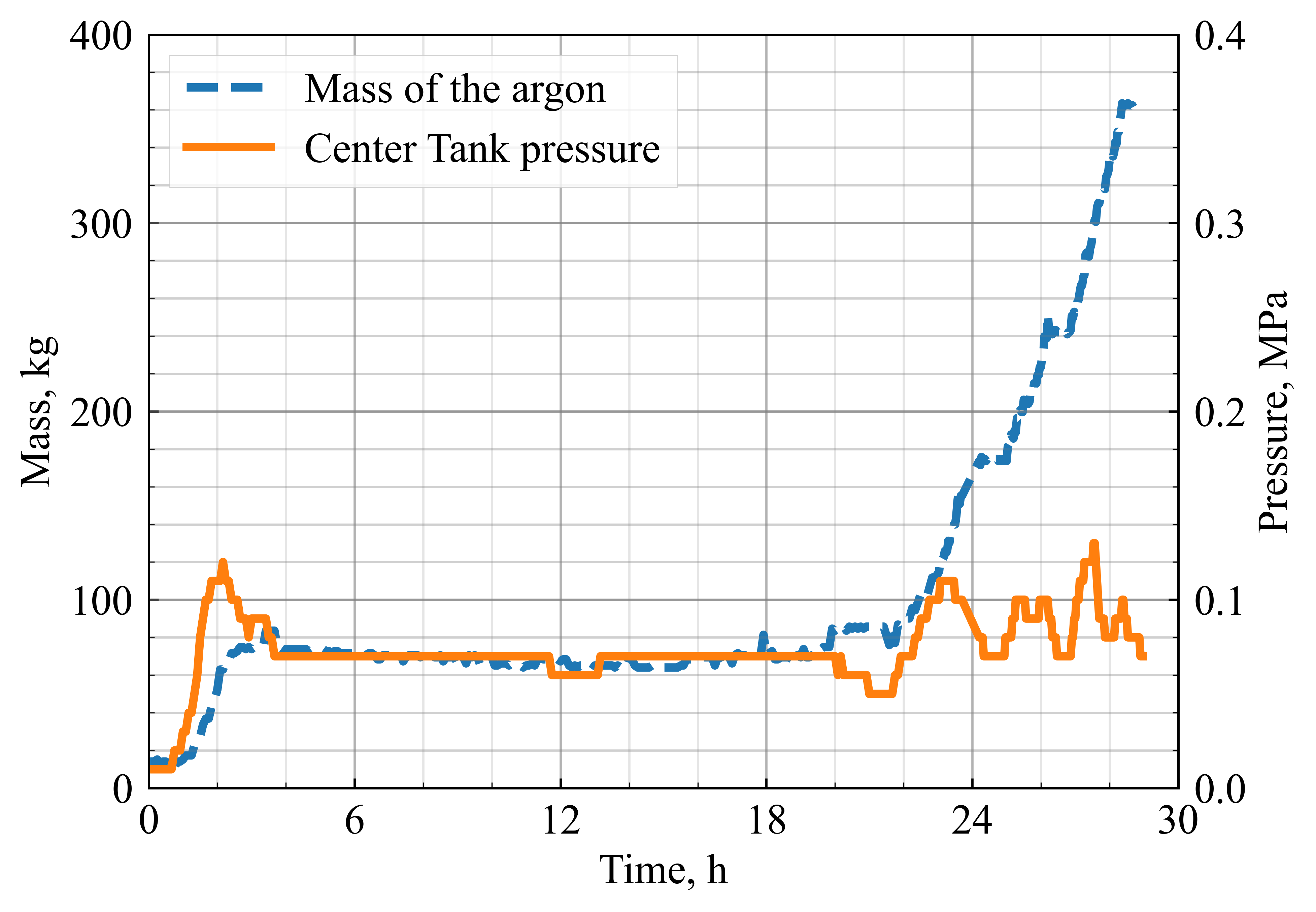}
  \caption{Argon mass, and Center Tank pressure during the liquefaction process.}
  \label{fig:liq}
  \end{figure}

After the Center Tank was completely cooled down, nearly 400 kg argon was liquified at the mass flow rate of 200 SLPM (\textasciitilde19 kg/h) to 900 SLPM (\textasciitilde87 kg/h) using the LN2 cooling circuit and the condenser. The mass of the argon inside the Center Tank as well as the Center Tank pressure during the liquefaction are presented in figure \ref{fig:liq}. During the liquefaction, the pressure was relatively stable at \textasciitilde0.1 MPa. The Center Tank pressure depends on the balance between the gas argon filling rate, the cooling power, and the inner vessel temperature. If the cooling power provided by the heat exchangers is larger than the cooling power needed for liquifying the GAr, the inner vessel temperature and Center tank pressure drops. Finally, the liquefaction rate reached 900 SLPM (\textasciitilde87 kg/h) with the operation of the LN2 cooling circuit and condenser, where the exhaust GN2 mass flow rate of the LN2 cooling circuit and the condenser were \textasciitilde49 kg/h and \textasciitilde64 kg/h, respectively.

To speed up the filling process, the Center Tank was filled LAr up to \textasciitilde3000 kg through the liquid inlet while the gas venting valve was kept open to perform vent filling rather than liquifying \textasciitilde3000 kg GAr. The utmost liquid level was approximately 1.31 m. Next, the liquid inlet valve was closed, and the gas venting valve was kept open for 48 hours to reach thermal equilibrium.

\subsubsection{Heat loss of the Center Tank} \label{sec:heat-loss}

After the LAr temperatures and the Center Tank pressure were stable, a mass flow meter (U.S.A. Alicate \cite{alicat}, M-50SLPM-D) was connected to the outlet of the gas venting valve to measure the discharged gas within 24 hours. According to the law of energy conservation, the heat losses is calculated as follow:
\begin{equation}
  Q_{a r}=\frac{m_{o u t} \Delta h+m_l c_{p, l}\left(T_{l, f}-T_{l, i}\right)+m_v c_{p, v}\left(T_{v, f}-T_{v, i}\right)}{\Delta t}
  \end{equation}
where \(Q_{ar}\) is the heat loss, J; \(m_{out}\) is the total mass of the discharged gas in 24 hours, kg; \(\Delta h\) is the specific enthalpy difference between the discharged gas and the liquid inside the Center Tank, J/kg; \(m_{l}\) as well as \(m_{v}\) are the mass of the liquid and gas inside the Center Tank, respectively, kg; \(c_{p,l}\) and \(c_{p,v}\) are the specific heat capacity of the liquid and gas inside the Center Tank, J/kg/K; \(T\) is the temperature of the argon, measured in K. The subscripts \(l\) and \(v\ \)represent LAr and GAr, respectively, and the subscripts \(f\) and \(i\) represent the states at the end of the experiment and at the beginning of the experiment, respectively. \(\Delta t\) is the time interval, s, which equals 24 hours. In this test, the mass of the discharged gas \(m_{out}\) was measured as 29.68 kg, corresponding to 90 W heat loss whilst \(\Delta h\) was calculated as 262.75 kJ using CoolProp \cite{bell_2014_pure}. Four PT100 temperature sensors installed inside the liquid phase at the height of 0 m, 0.95 m, 1.1 m, and 1.25 m registered a gradual increase, with an average of 1.09 K, resulting in a heat loss of 41 W whilst the liquid specific heat capacity \(c_{p,l}\) equals 1117.20 J/kg/K. The energy change in the gaseous phase is negligible due to small mass of \(m_{v}\). As a result, the heat loss of the Center Tank is calculated as 131 W in total. According to the Chinese standard GB/T 18442-2019: Static vacuum insulated cryogenic pressure vessel, the evaporation rate of a commercial cryogenic vessel with a volume of 6 m$^3$, when 95\% of its volume is filled with saturated LAr at 0.1 MPa, should be less than 0.3\% per day in mass, corresponding to a heat loss rate of approximately 45 W. The heat loss of the Center Tank is greater than that of a commercial cryogenic vessel of the same volume. The reason is that the neck of the Center Tank has a wall thickness of 16 mm, which significantly increases the heat loss due to heat conduction. In this test, the heat loss through the neck is calculated as 66 W by Fourier's heat conduction law whilst the temperature at the bottom of the neck ($h=3.66$ m) and the ambient temperature equaled 155 K and 303 K, respectively, corresponding to 50\% of the total heat loss. Assuming the ratio of heat loss caused by heat conduction and radiation is 0.6 and 0.4, respectively, the heat losses of the Center Tank with LXe at the ambient temperature of 293 K can be calculated as follow:
\begin{equation}
  Q_{xe}=Q_{ar}\left(0.6 \times \frac{T_{sd}-T_{xe}}{T_{en}-T_{ar}}+0.4 \times \frac{T_{sd}^4-T_{xe}^4}{T_{en}^4-T_{ar}^4}\right)
  \label{eq:heatlossxe}
  \end{equation}
where \(T_{sd}\) is the temperature at the standard condition, 293 K; \(T_{xe}\) and \(T_{ar}\) are the saturation temperature of LXe and LAr, which equals 165 K and 87 K, respectively; \(T_{en}\) is the ambient temperature, 303 K; By applying eq. (\ref{eq:heatlossxe}), the heat losses of the Center Tank is predicted as 89 W at the ambient temperature of 293.15 K with 35\% of its volume filled with LXe at 165 K. Therefore, \textasciitilde180 L liquid nitrogen per day is required to ensure stable storage of 30 tons xenon (assuming the outlet temperature of the LN2 cooling module is 150 K).

\subsection{Pressure regulation using the heat exchangers} \label{sec:exp_pre_reg}

After testing the heat losses of the Center Tank, the Center Tank was sealed for ten days. Subsequently, the Center Tank was depressurized by the LN2 cooling circuit, then pressurized by the heater, and finally depressurized again by the condenser.

\begin{figure}[H]
  \centering
  \subfloat[]{
  \begin{minipage}[c]{0.49\textwidth}
  \centering
  \centerline{\includegraphics[width=1\textwidth]{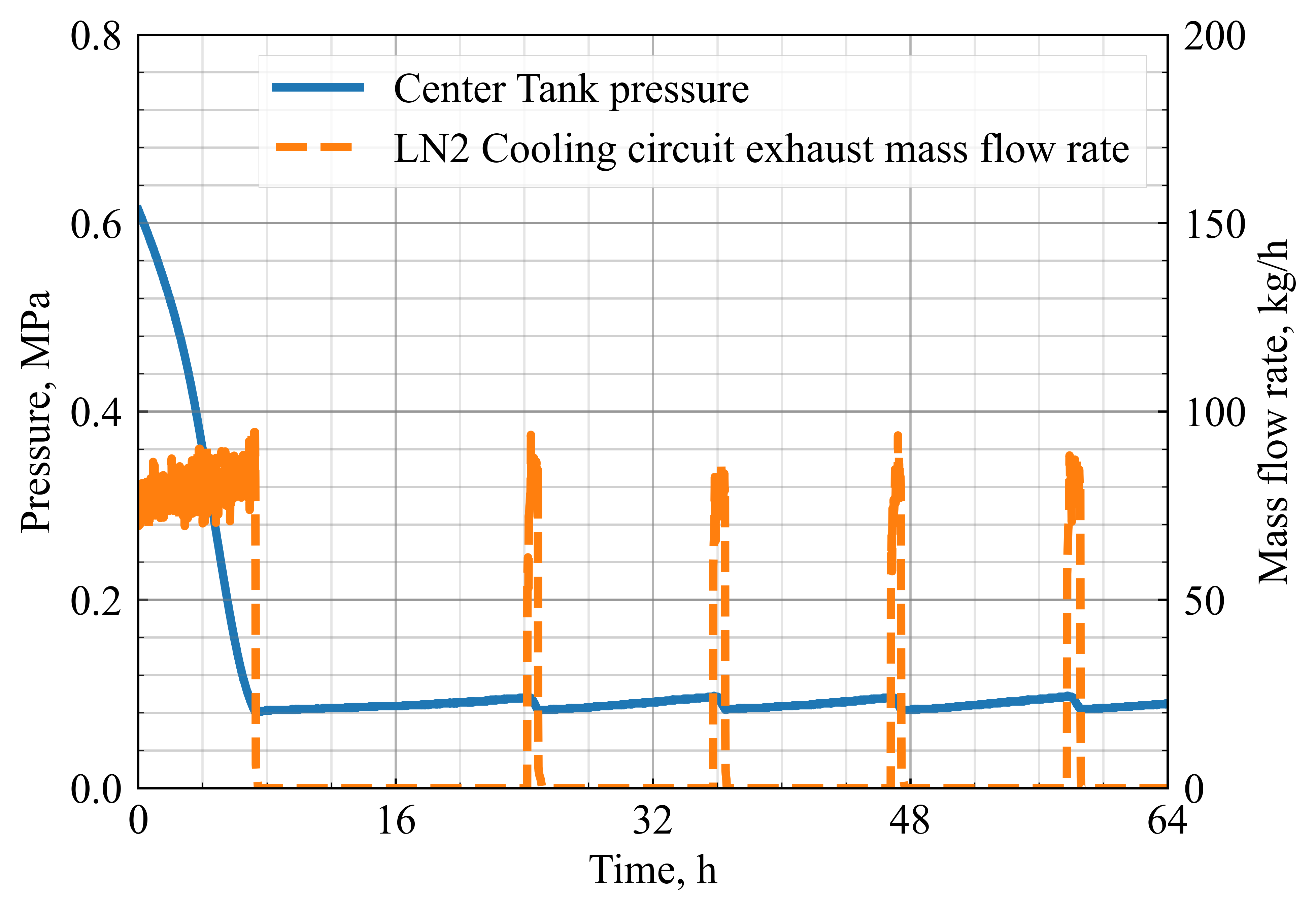}}
  \end{minipage}
  }
  \centering
  \subfloat[]{
  \begin{minipage}[c]{0.49\textwidth}
  \centering
  \centerline{\includegraphics[width=1\textwidth]{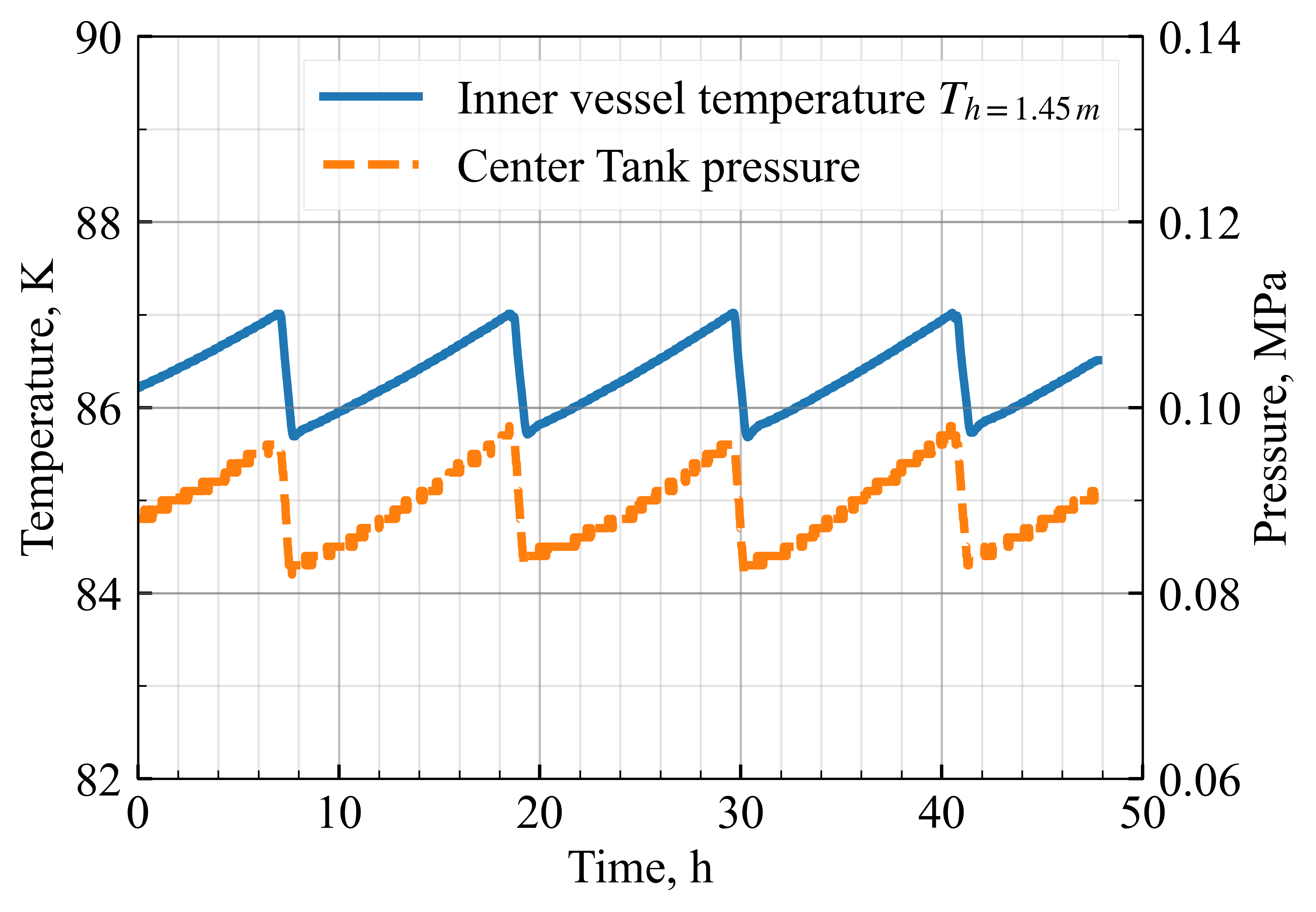}}
  \end{minipage}
  }
  \caption{(a). The Center Tank pressure and exhaust GN2 mass
  flow rate of the LN2 cooling circuit during the operation of cooling
  circuit; (b). $T_{h=1.45\, \rm{m}}$ and the Center Tank pressure in the
  stable storage process.}
  \label{fig:pre_wall}
  \end{figure}

Before operating the LN2 cooling circuit, the control strategy described in Section \ref{sec:heat_exchanger} was adopted with the trigger temperatures set as 86 K and 87 K, aiming to stabilize the Center Tank pressure at approximately 0.1 MPa. The Center Tank pressure, as well as the exhaust GN2 mass flow rate of the LN2 cooling circuit, are displayed in figure \ref{fig:pre_wall}(a). After the inlet valve of the cooling circuit was turned on, the Center Tank pressure dropped rapidly from 0.6 MPa to 0.1 MPa, and then remained at \textasciitilde0.1 MPa after $T_{h=1.45\, \rm{m}}$ decreased to the lower limit of the trigger temperature.

During the stable storage process, the LN2 cooling circuit was operated for 40 minutes in every 12 hours to compensate for the heat losses. $T_{h=1.45\, \rm{m}}$ and the Center Tank pressure during the stable storage process, are further displayed in figure \ref{fig:pre_wall}(b). The lowest temperature of the $T_{h=1.45\, \rm{m}}$ was 0.3 K lower than the lower trigger temperature of 86 K, which was due to the residual LN2 inside the cooling circuit after the inlet valve was closed. As shown in figure \ref{fig:pre_wall}(b), the Center Tank pressure was stabilized at 0.09\(\pm\)0.008 MPa, and the LAr temperature at $h =0$ m was stabilized between 85.4 to 86.3 K during the stable storage process. The $P_{\rm{Center \, Tank}}$ and $T_{h=1.45\, \rm{m}}$ have the same trend, thereby the $P_{\rm{Center \, Tank}}$ can be controlled as intended by setting the trigger temperatures appropriately.

\begin{figure}[H]
  \centering
  \includegraphics[width=0.6\textwidth]{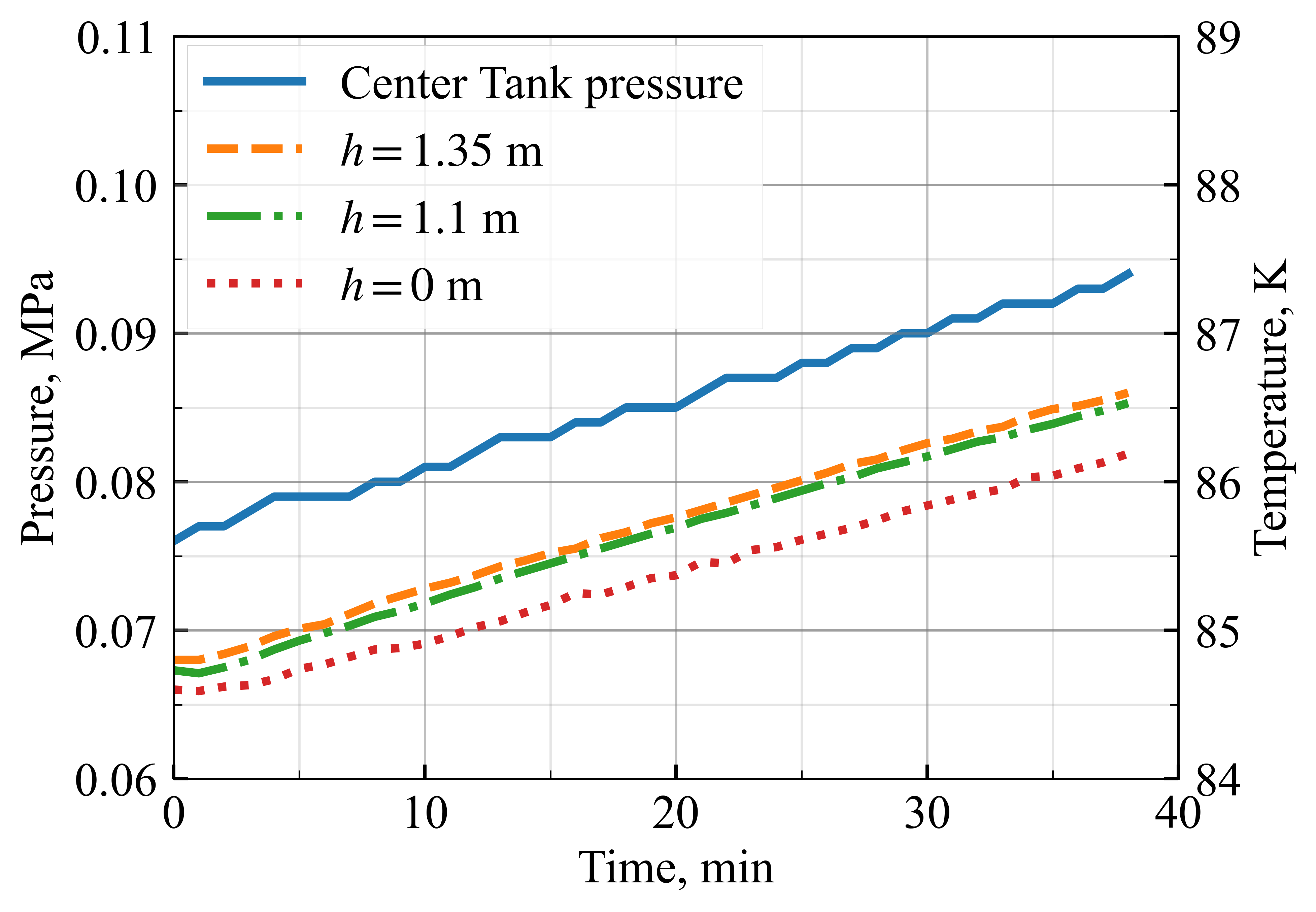}
  \caption{The Center Tank pressure and liquid argon temperatures at $h =0$ m, 1.1 m, 1.25 m during the operation of the heater.}
  \label{fig:heater}
  \end{figure}

During the operation of the GN2-based heater, the pressure of the Center Tank, as well as the temperature of the LAr, increased synchronously, as shown in figure \ref{fig:heater}. The Center Tank pressure increased from 0.076 MPa to 0.093 MPa in approximately 40 minutes whilst the average heating power is calculated as 4000 W. The LAr temperature at different heights increased with the same trend, indicating that the heat was transferred evenly to the entire liquid region.

\begin{figure}[htpb]
  \centering
  \includegraphics[width=0.6\textwidth]{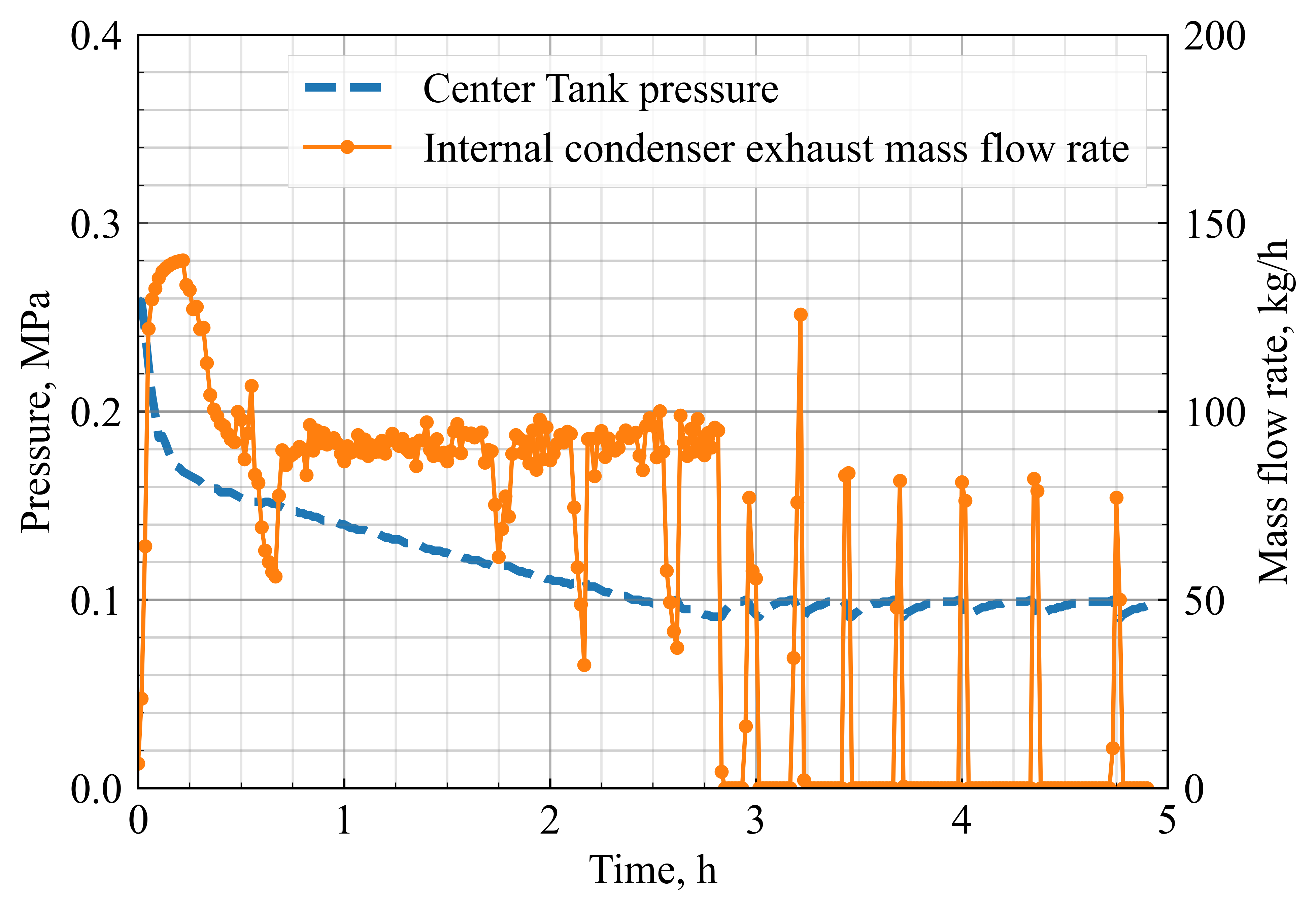}
  \caption{The Center Tank pressure and exhaust GN2 mass flow rate of the condenser during the operation of the condenser.}
  \label{fig:con}
  \end{figure}

After the heater was turned off and the LAr temperature became stable, the condenser depressurized the Center Tank again. The control strategy described in Section \ref{sec:heat_exchanger} was adopted and the trigger pressures were set as 0.09 MPa and 0.1 MPa. As shown in figure \ref{fig:con}, after the condenser was turned on, the pressure of the Center Tank drops from 0.2 MPa to 0.1 MPa, and then stabilizes between 0.09 MPa and 0.1 MPa. However, the exhaust GN2 mass flow rate of the condenser was unstable due to the small pipe diameter (16 mm) and structure---three sets of coils connected in parallel.

The condenser and the cooling circuit depressurize the Center Tank pressure differently. In the experiment, the condenser and LN2 cooling circuit operated at similar mass flow rates for 5 minutes and 10 minutes, respectively, to depressurize the Center Tank by 0.01 MPa. Meanwhile, the temperature of $T_{h=1.45\, \rm{m}}$ deceased by 0.1 K and 1.0 K, respectively, indicating the cooling capacity provided by the condenser was used to reducing the Center Tank pressure, while most of the cooling capacity provided by the cooling circuit was used to cooling the inner vessel. Therefore, the condenser can quickly depressurize the Center Tank while the cooling circuit enables the Center Tank inner vessel to act as a thermal buffer, storing and then providing cooling power. In summary, three heat exchangers can regulate the Center Tank pressure effectively as designed.

\subsection{Non-vented filling and recovery} \label{sec:exp_nvf_nvr}
\subsubsection{Non-vented liquid filling (NVF)} \label{sec:exp_nvf}

In the NVF experiments, the ideal final pressure and the MAWP of the Test Detector is assumed to be 0.2 MPa and 0.3 MPa, respectively. Two strategies mentioned in Section \ref{sec:nvf} were adopted to avoid an overpressure of the Test Detector, including reducing the LAr temperature in the Center Tank and precooling the Test Detector before NVF. For the Center Tank, the LAr temperature was reduced and maintained by the LN2 cooling circuit with the trigger temperatures of 86 K and 87 K. For the Test Detector, the inner vessel was cooled by the LN2 cooling circuit with the trigger temperatures of 86 K (2 K above the argon freezing point to avoid argon freezing) and 87 K. In the NVF experiment, as $T_{h=1.45\, \rm{m}}$ of the Test Detector decreased to 86 K, the inner vessel of the Test Detector that was intended to be in contact with the LAr was completely cooled down.

\begin{figure}
  \centering
  \includegraphics[width=0.6\textwidth]{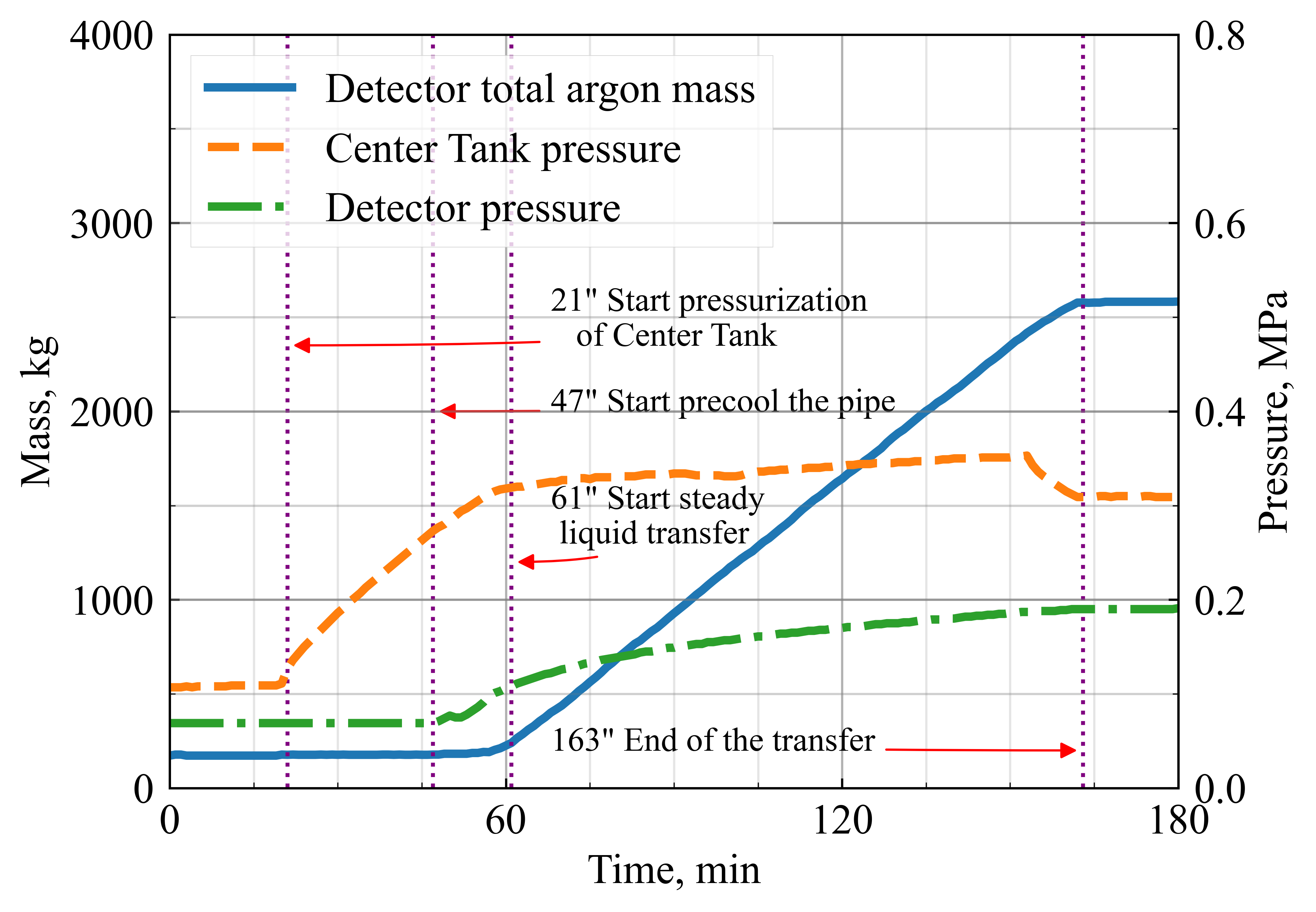}
  \caption{The argon mass inside the Test Detector, the pressure of the Center Tank and Test Detector during the NVF process.}
  \label{fig:nvf}
  \end{figure}

After the LAr in the Center Tank was maintained at approximately 86 K (LAr temperature at $h =0$ m) and the Test Detector was fully cooled down, pressurization of the Center Tank was followed by injecting GAr into the ullage. The control range of the Center Tank pressure was set as 0.3 MPa to 0.35 MPa to overcome the flow resistance between the Center Tank and the Test Detector. As shown in figure \ref{fig:nvf}, the Center Tank pressure increased to 0.28 MPa in approximately 30 minutes. Then, all the valves in the liquid transfer line between the Center Tank and the Test Detector opened to start the filling.

At the beginning of the NVF, the \textasciitilde26 m long transfer line was at ambient temperature. As the LAr was driven to the Test Detector through the transfer line, it evaporated due to the heat transfer between the LAr and the pipeline. Hence, the argon was transferred in the gaseous phase to the Test Detector. After the pipeline was completely cooled down, in \(\sim\)10 minutes, the argon was transferred in the liquid phase at an average mass flow rate of \(\sim\)1390 kg/h, measured by the Coriolis mass flow meter (Emerson \cite{emerson} CMF100) installed in the middle of the pipeline. Meanwhile, the argon mass inside the Test Detector increased rapidly during the liquid transfer period, as shown figure \ref{fig:nvf}. The mass flow rate of LXe is 2140 kg/h with the same Reynolds number. From the beginning to the end of the filling, the Test Detector pressure continuously increased and reached 0.2 MPa. The increase in the pressure of the Test Detector was mainly caused by the compression of the gaseous region as the liquid entering the Test Detector, the heat losses along the pipeline, and the heat transfer between the gas argon and the upper part of the Test Detector. It should be noted that only two of the three strategies mentioned in the Section \ref{sec:nvf} were applied to reduce the Test Detector pressure, and the Test Detector pressure can be further decreased by operating the cooling modules coupled with the Test Detector.

\subsubsection{Non-vented liquid recovery (NVR)} \label{sec:exp_nvr}

Before the NVR, the LAr in the Test Detector remained static at about 93.6 K ($h =0$ m) by operating the condenser with the trigger pressures of 0.18 MPa and 0.2 MPa. The inner vessel of the Center Tank was intentionally precooled to a high temperature of 100 K ($T_{h=1.45\, \rm{m}}$) instead of 86 K, in order to test the depressurization performance of the condenser during the NVR process whilst the trigger pressures of the condenser were set at 0.18 MPa and 0.2 MPa to save LN2.

The NVR in the preliminary test mainly consisted of three stages. Firstly, the Test Detector was pressurized by injecting GAr from the gas bottles, while the pressurization pressure was set at 0.4 MPa and 0.45 MPa. After the Test Detector pressure reached 0.35 MPa, the valves of the liquid recovery line were opened, and the LAr inside the Test Detector was pushed to the cryogenic pump via the liquid recovery line by the pressure difference between the Test Detector and the Center Tank. The pipeline was precooled by the LAr. Secondly, when the cryogenic pump was completely cooled, and the argon at the inlet of the cryogenic pump was in the liquid phase, the cryogenic pump was turned on with a frequency of 60 Hz. In order to simulate the pressure drop caused by height difference, the opening of the control valve was adjusted to 43\% before the pump was turned on. The pressure drop of the control valve was approximately 0.28 MPa, corresponding to a head of a 20.5 m LAr column. Finally, with most of the LAr inside the Test Detector recovered to the Center Tank, the cryogenic pump was turned off, and the opening of the control valve was set to 100\%. The NVR experiment focuses on the second stage, where the recovery is driven by the cryogenic pump. It should be noted that in CJPL, as the recovery process starts, the liquid inside the detector flows to the cryogenic pump by gravity without the need to pressurize the detector.

\begin{figure}
  \centering
  \includegraphics[width=0.6\textwidth]{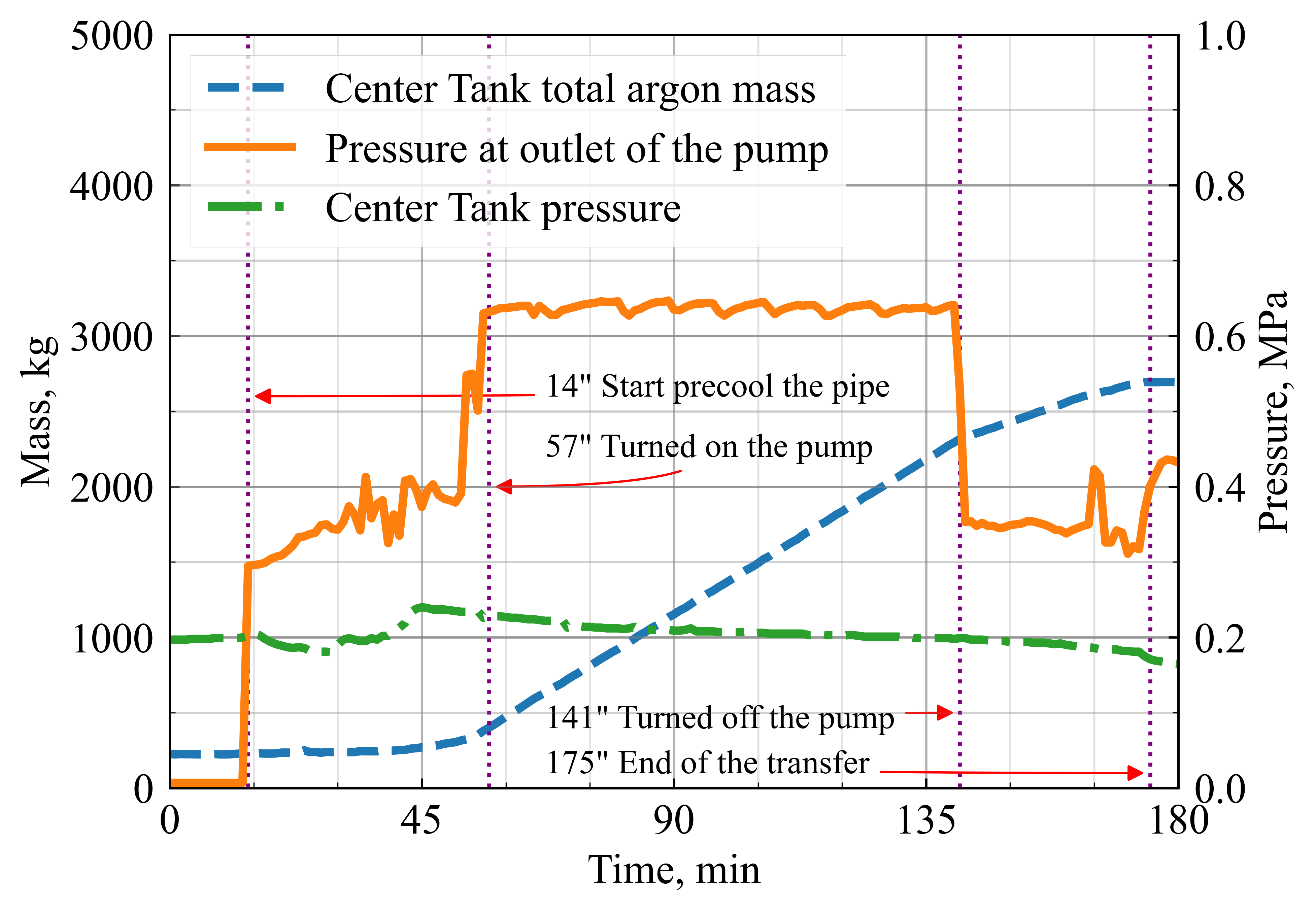}
  \caption{The pressure of the Center Tank and at the outlet of the pump, and the argon mass inside the Center tank during the NVR process.}
  \label{fig:nvr}
  \end{figure}

In the second stage of NVR, after the pump was turned on, the pressure at the outlet of the cryogenic pump increased rapidly from 0.3 MPa to 0.63 MPa, as shown in figure \ref{fig:nvr}. The average mass flow rate was 1395 kg/h, and the pressure difference between the inlet and the outlet of the pump was about 0.2 MPa, corresponding to the pressure a 15.8 m LAr column. Meanwhile, the pressure of the Center Tank was reduced by the condenser with a cooling power of 8000 W. At the end of the NVR, a total mass of \textasciitilde2500 kg argon was recovered and only \textasciitilde360 kg of LN2 was used to control the Center Tank pressure.

\section{Conclusion}\label{sec:conclusion}

A novel full-scale cryogenic liquid xenon handling facility, the First-X, was designed, constructed, and tested for the PandaX-30T, the next-generation of the dark matter and neutrinos experiment. The First-X is designed to liquefy and store xenon long-term without losses and contamination, to fill the detector with cryogenic liquid xenon and recover cryogenic liquid xenon from the detector to the storage module safely and effectively without venting out. The storage module of the First-X is five Center Tanks, which are designed with a vacuum and multi-layer insulation and a maximum allowable working pressure of 7.1 MPa, allowing 6 tons of xenon to be stored at 165--178 K at 0.1--0.2 MPa in the liquid phase or up to 300 K and up to 6.95 MPa in the supercritical phase. The heat loss is compensated by the cooling modules whilst xenon is stored at 165--178 K at 0.1--0.2 MPa. High-pressure storage (\textgreater0.2 MPa) only occurs in case of long-term detector shutdown or lack of nitrogen, ensuring no-loss storage of 6 tons of xenon in the range 178--300 K. The inner surface roughness of the pipeline and the storage module is below 0.4 \(\mu\)m to reduce the outgassing contamination, and the helium leak rate of all the seals is lower than 1\(\times\)10\(^{- 10}\) Pa\(\cdot\)m\(^{3}\)/s to reduce the air leak contamination.

According to the experiment, the heat loss of the Center Tank with 6 tons of xenon stored at \textasciitilde165 K is estimated to be 89 W at an ambient temperature of 293 K with a liquid filling ratio of 35\%. Therefore, \textasciitilde180 L liquid nitrogen per day is required to ensure stable storage of 30 tons xenon (assuming the outlet temperature of the LN2 cooling module is 150 K). The pressure of the Center Tank can be regulated effectively within a range of 0.085\(\pm\)0.005 MPa. A non-vented filling (NVF) and non-vented recovery (NVR) of cryogenic LAr were conducted successfully at the mass flow rate of approximately 1390 kg/h, corresponding to a projected liquid xenon mass flow rate of 2140 kg/h. Compared with the PandaX-4T, the transportation of xenon between the detector and the storage module is conducted in the liquid phase rather than in the gaseous phase, and the filling rate (fill the detector) and the recovery rate (recover xenon from the detector) are increased by approximately 50 times and 30 times, respectively.

\acknowledgments
This project is supported in part by a grant from the National Science Foundation of China (Nos. 12090061). We thank the Tsung-Dao Lee Institute Experimental Platform Development Fund for their support. We also thank Hongwen Foundation in Hong Kong, and Tencent Foundation in China for their sponsorship. Finally, we thank the Chengdu Xinliantong Cryogenic Equipment Co.,Ltd for their construction work and other assistance.



\end{document}